\documentclass[
    reprint,
    superscriptaddress,
    amsmath,
    amssymb,
    aps,
    physrev,
    nofootinbib
]{revtex4-2} 

\usepackage{graphicx}
\usepackage{dcolumn}
\usepackage{bm}
\usepackage{xcolor}
\usepackage{hyperref}
\usepackage{amsmath}
\usepackage{booktabs}
\usepackage{siunitx}
\usepackage{cancel}
\usepackage{ulem}
\usepackage{newtxtext,newtxmath}
\usepackage{multirow}
\usepackage{mathtools}

\def\placeholder{{\square}}

\definecolor{mypaleblue}{RGB}{44,45,142}

\begin{document}

\preprint{APS/123-QED}

\title{\textbf{Modeling the CMB Optical Depth from Reionization Physics: A Symbolic Regression Approach} 
}

\author{Marek Detière-Venkatesh}
 \email{Corresponding author: marek.detiere-venkatesh@mail.mcgill.ca}
\author{Adrian Liu}
\author{Debanjan Sarkar}
\affiliation{Department of Physics \& Trottier Space Institute, McGill University, Montréal, QC H3A 2T8, Canada}
\affiliation{Ciela - Montreal Institute for Astrophysical Data Analysis and Machine Learning}

\author{Wassim Tenachi}
\affiliation{Ciela - Montreal Institute for Astrophysical Data Analysis and Machine Learning}
\affiliation{Universit\'e de Strasbourg, CNRS, Observatoire astronomique de Strasbourg, UMR 7550, F-67000 Strasbourg, France}
\affiliation{Mila - Quebec Artificial Intelligence Institute}

\date{\today}

\begin{abstract}
The cosmic microwave background optical depth $\tau$ is a key summary parameter for understanding reionization physics. The dependence of $\tau$ on other cosmological and astrophysical parameters is often simulated using numerical codes, which are slow to evaluate. Here we explore the use of symbolic regression, specifically the automated equation discovery algorithm \texttt{PySR}, for emulating reionization codes. Testing on \texttt{21cmFAST}, \texttt{AMBER}, and \texttt{Zeus21}, we find emulation errors $\Delta\tau$ to be less than $10^{-3}$. Symbolic regression also enables a reparametrization of some codes in terms of the parameters of others, allowing for a fair comparison between $\tau$ predictions from different codes. As a test application, we perform a Fisher forecast for combined parameter constraints from 21 cm, CMB, and galaxy surveys, paying particular attention to systematic modeling differences in $\tau$ predictions arising from the use of the three aforementioned codes and finding meaningful differences between the codes that can cause biases in the parameter predictions. Overall, this exercise showcases the utility of symbolic regression for comparing theory models.

\end{abstract}

\maketitle

\section{Introduction}

The Epoch of Reionization (EoR) is a crucial period in the universe's history, representing the time when the first galaxies emitted UV radiation that ionized the surrounding neutral hydrogen in the intergalactic medium
\cite{Barkana:2000fd,Furlanetto:2006jb,Fan:2006dp,Robertson:2021ljt,Gnedin:2022eza}. This effect is imprinted in the cosmic microwave background (CMB) optical depth $\tau$, a scalar quantifying the probability a CMB photon would scatter off a free electron released during reionization \cite{Zaldarriaga:1996ke,Barkana:2000fd,Planck:2018vyg}. It can therefore be expressed in terms of the sky-averaged electron density history $\bar n_e(z)$ through
\begin{equation}
    \tau = \sigma_T\int \bar n_e(z) \frac{dl}{dz}dz,
    \label{eq:tau}
\end{equation}
where $\sigma_T$ is the Thomson cross-section, and $dl/dz$ is the proper line-of-sight distance element per unit redshift. The optical depth is often also considered a nuisance parameter in CMB studies given its degeneracies with fundamental cosmological parameters such as the scalar amplitude of primordial fluctuations $A_{\rm s}$. Independent probes of $\tau$ can therefore complement CMB measurements by breaking these degeneracies, thereby improving cosmological parameter constraints \cite{Liu:2015txa,Shmueli:2023box}. 

However, $\tau$ typically cannot be constrained from observations alone, but requires the use of cosmological simulations. These simulations are often computationally challenging because they require resolving complex small-scale astrophysical processes that collectively shape the large-scale distribution of neutral hydrogen
\cite{Gnedin:2022eza,Trac:2009bt,Iliev:2013una,Kannan:2021xoz}. This can be problematic when simulations need to be run repeatedly for inference exercises. Furthermore, it is not even clear that the inference results will necessarily be accurate since simulations may disagree on their physics prescriptions and numerical implementations, and there does not exist an agreed-upon ``ground truth'' for reionization predictions.

In order to circumvent the time-intensive nature of such codes, neural networks are often utilized to emulate simulation outputs \cite{Kern:2017ccn,Schmit:2017pho,Jennings:2018eko,Breitman:2023pcj,Lazare:2023jkg,Jense:2024llt,Breitman:2026fte}. While these emulators are fast and precise, they have the downside of functioning as ``black boxes" in that they do not inform us about how model parameters influence the simulated quantity. They cannot describe the mathematical mappings between simulations' inputs and outputs explicitly, making it difficult to extract meaningful physics from the emulators; for example, a neural network can accurately predict outputs without highlighting parameter degeneracies. This lack of transparency hinders our ability to use them as a tool to understand the differences between simulations. A promising alternative to these opaque machine-learning methods is symbolic regression (SR), a more interpretable approach for parsing complex astrophysical datasets into transparent, analytic expressions.

SR is the automated discovery of equations that map physical observables to relevant input variables. Currently, various packages implement SR using diverse methodologies, such as deep reinforcement learning \cite{Tenachi2023, Tenachi2024}, physically motivated simplifying algorithms \cite{Udrescu2020}, exhaustive searches \cite{Bartlett2023ESR}, and genetic algorithms \cite{Cranmer2023, Eureqa}. Unlike other machine-learning methods, the equations returned by SR can concisely summarize the complex physics underlying these simulations, making them more interpretable. Furthermore, in some applications, analytic models obtained through SR exhibit superior extrapolation capabilities compared to their neural network counterparts \cite{EQL_SR, Kamienny_SR_for_RL}. These expressions are also often less likely to overfit to data. By modeling multiple simulations in this analytic manner, we can perform direct comparisons of their underlying parameter dependencies. This makes SR a valuable tool for directly probing similarities and discrepancies between different simulations of reionization theory, effectively connecting physical parameters of interest to observable quantities. Related applications include Ref. \cite{SR_Omegam_from_galphoto}, where a formula relating the matter density parameter $\Omega_{\rm m}$ to galactic photometric properties was successfully regressed from simulation data, and Ref. \cite{Bartlett:2023cyr}, where precise formulae relating the scalar amplitude $A_{\rm s}$ and the linear matter power spectrum to other cosmological parameters were produced.

In this work, we use symbolic regression to obtain precise analytic formulae for $\tau$ regressed on data from three cosmological simulations: \texttt{21cmFAST} \cite{Mesinger2011,Murray:2020trn}, \texttt{AMBER} \cite{Trac2022}, and \texttt{Zeus21} \cite{Munoz2023,Cruz:2024fsv}. This represents one of the first applications of SR to reionization physics. Our equations are accurate and can be partially interpreted, providing insight into the inner workings of these simulations. Furthermore, we utilize SR to re-express the input variables of the simulations within a common basis, enabling a direct comparison of the behavior of $\tau$ across different codes. This framework offers a novel approach to inter-code comparison and, importantly, can be extended to simulations beyond those considered in this work. 

We also explore the use of our analytic equations in forecasting cosmological parameters constraints, with a particular emphasis on testing the consistency between the different simulations in this exercise. As a concrete worked example, we update the Fisher forecasts for combining 21 cm observations, the CMB, and galaxy surveys that were detailed in Ref.~\cite{Liu:2015txa}. Our update encompasses both methodology (replacing numerical derivatives with SR-derived expressions) and observational specifications (making use of more recent experiments). Since we can easily perform these forecasts for each of the three codes on a common basis, our results are a test of robustness to different theory model choices.

We organize the paper as follows: in Section~\ref{sec:methods}, we detail the theory of the optical depth $\tau$ and introduce the three studied cosmological simulations and the symbolic regression algorithm we employ. In Section~\ref{sec:results}, we study the accuracy of our symbolic regression-derived expressions for $\tau$ and physically interpret them where possible. In Section~\ref{sec:simulation_comparison}, we use SR to link the astrophysical parameters between our simulations and express our $\tau$ formulae in a common parameter basis, enabling us to compare the simulations. Finally, we build a Fisher matrix formalism to improve forecasts on cosmological parameters in Section~\ref{sec:fisher_forecast}. We summarize our conclusions in Section~\ref{sec:summary_discussion}.

\section{Methods}\label{sec:methods} 

\subsection{Cosmological simulations}
\label{subsec:meth:sims}

Here, we detail the simulations used to compute $\bar n_e(z)$, and hence $\tau$ using Eq.~\eqref{eq:tau}. These are parametrized by a set of cosmological and astrophysical parameters, which we vary in order to produce a diverse set of training scenarios for SR to fit to. The parameters that we chose to vary, along with their corresponding ranges, are presented in Tables~\ref{tab:cosmo_params} and \ref{tab:astro_params} for cosmological and astrophysical parameters, respectively.

\begin{table}[t]
\setlength{\tabcolsep}{6pt}
\centering
\small
\caption{Cosmological parameters varied in the simulations. Each use a set of 5 parameters which encode their cosmology. The chosen ranges correspond to $\pm3$ standard deviations from the fiducial cosmological parameter values \cite{Planck:2018vyg}.}
\label{tab:cosmo_params}
\renewcommand{\arraystretch}{1.15}
\begin{tabular}{l l l l}

\toprule
Parameter & Used in & Min & Max \\
\midrule

$h$ & All & 0.659 & 0.689 \\
$n_{\rm s}$ & All & 0.953 & 0.977 \\
$\Omega_{\rm m}$ & \texttt{21cmFAST}, \texttt{AMBER} & 0.294 & 0.336 \\
$\Omega_{\rm b}$ & \texttt{21cmFAST}, \texttt{AMBER} & 0.0470 & 0.0516 \\
$\sigma_8$ & \texttt{21cmFAST}, \texttt{AMBER} & 0.793 & 0.829 \\
$\Omega_{\rm c} h^2$ & \texttt{Zeus21} & 0.117 & 0.123 \\
$\Omega_{\rm b} h^2$ & \texttt{Zeus21} & 0.0221 & 0.0227 \\
$A_{\rm s}$ & \texttt{Zeus21} & $2.00\times10^{-9}$ & $2.19\times10^{-9}$ \\

\bottomrule
\end{tabular}
\end{table}

\begin{table*}
\setlength{\tabcolsep}{6pt}
\centering
\small
\caption{Astrophysical parameters varied in the simulations. Parameter definitions in simulations' analytic equations are available in \cite{Park:2018ljd} for \texttt{21cmFAST}, \cite{Trac2022} for \texttt{AMBER}, and \cite{Munoz2023} for \texttt{Zeus21}, with \texttt{Zeus21}'s $f_{\star, 10}$ serving as a proxy for the native $\epsilon_\star$ parameter as described in Sec. \ref{subsec:meth:sims}. The chosen ranges for the first five astrophysical parameters correspond to $\pm1$ standard deviations from the HERA Phase I limits \cite{HERA_2022}.} 
\label{tab:astro_params}
\renewcommand{\arraystretch}{1.15}
\resizebox{\textwidth}{!}{
\begin{tabular}{l l l c c}

\toprule
Parameter & Used in & Description & Min & Max \\
\midrule

$f_{\star,10}$ & \texttt{21cmFAST, Zeus21} & Fraction of galactic gas in stars for $10^{10}M_{\odot}$ halos, in $\log_{10}$ units & -1.40 & -1.04 \\
$\alpha_\star$ & \texttt{21cmFAST, Zeus21} & Fraction of galactic gas in stars power index & 0.41 & 0.62 \\
$f_{\rm esc,10}$ & \texttt{21cmFAST, Zeus21} & Escape fraction of ionizing photons for $10^{10}M_{\odot}$ halos, in $\log_{10}$ units & -1.00 & -0.52 \\
$\alpha_{\rm esc}$ & \texttt{21cmFAST, Zeus21} & Escape fraction of ionizing photons power-law index & -0.52 & 0.34 \\
$M_{\rm turn}$ & \texttt{21cmFAST, Zeus21} & Turnover mass below star formation is suppressed, in $\log_{10}$ solar mass units & 8.20 & 9.17 \\ 
$R_\text{bubble}^\text{max}$ & \texttt{21cmFAST} & Ionizing photons' mean free path within ionizing bubbles, in Mpc & 10 & 30 \\
$z_\text{mid}$ &\texttt{AMBER} &Midpoint of reionization & 6.0 & 11.0 \\
$\Delta_z$ &\texttt{AMBER}& Duration of reionization & 0.1 & 7.0 \\
$A_z$ &\texttt{AMBER} & Asymmetry of reionization & 1.0 & 5.0 \\

\bottomrule
\end{tabular}
}
\end{table*}

The first simulation we consider is \texttt{21cmFAST} \cite{Mesinger2011}\footnote{\href{https://github.com/21cmfast/21cmFAST.git}{https://github.com/21cmfast/21cmFAST.git}}, an excursion-set based code for quickly computing ionization fields during reionization, in addition to including effects such as X-ray heating. The code is parametrized by astrophysical parameters encoding structure formation and radiation flux (presented in Table \ref{tab:astro_params}). Among other products, the simulation outputs the mean neutral fraction of hydrogen $\bar x_\text{HI}$ and assumes $\textrm{He II}$ ionization happens at $z=3$, allowing it to compute $\tau$ with Eq.~\eqref{eq:tau} by expanding $\bar n_e(z)$ into its hydrogen and helium contributions. Our simulations are run on (256 Mpc)$^3$ boxes with a spatial resolution of 1 Mpc and on a redshift grid of resolution $\Delta z=0.05$ for $5\le z\le 25$, and $\Delta z = 0.5$ for $25 \le z \le 50$. We generate 2048 realizations of \texttt{21cmFAST} with parameters sampled by a Sobol sequence \cite{sobol_1967} within the ranges presented in Tables~\ref{tab:cosmo_params} and \ref{tab:astro_params}. We retain 2025 of these, as 23 simulations had not fully ionized by $z=5.0$, making them unlikely to be consistent with current observations even with a reasonably generous interpretation of data on how late reionization may have taken place.
We also keep $t_\star$, the star-formation timescale,
and $L_X$, the X-ray luminosity normalization, fixed
at their default values. Our exploratory calculations
indicated that their effects on $\tau$ are negligible
for the configurations considered here. This is
consistent with the ionization history in \texttt{21cmFAST} 
being primarily determined by the cumulative production
of ionizing UV photons, rather than the instantaneous
star-formation rate or X-ray heating.

The second simulation we consider is \texttt{AMBER}~\cite{Trac2022} \footnote{\href{https://github.com/hytrac/amber.git}{https://github.com/hytrac/amber.git}}, which is a semi-numerical code catered towards reionization studies \cite{Trac2022}. Given an ionization history, this simulation determines the order of individual cell ionization by ranking them in terms of radiation intensity to build a reionization redshift field \cite{Chen:2022lhr}. Similarly to \texttt{21cmFAST}, \texttt{AMBER} computes the ionization history $\bar n_e$ to produce $\tau$. However, the astrophysical parameters it employs are  phenomenological descriptions of reionization (see Table~\ref{tab:astro_params}), granting the user greater control over the ionization history. We generate 1024 realizations of \texttt{AMBER} with the same sampling strategy detailed previously. They are generated with (500 Mpc/h)$^3$ boxes with 512 cells per side. Unused cosmological and astrophysical parameters are fixed to their fiducial values. We omit the reionization parameters $M_\text{h}$ and $\lambda_\text{mfp}$ as we found they did not impact output $\tau$ values.

Finally, we include \texttt{Zeus21} \cite{Munoz2023} \footnote{\href{https://github.com/ZeusCosmo/Zeus21.git}{https://github.com/ZeusCosmo/Zeus21.git}}, a third, markedly different simulation of Cosmic Dawn. \texttt{Zeus21} uses a semi-numerical technique for modeling the thermal and ionization evolution of the intergalactic medium during Cosmic Dawn and the Epoch of Reionization, while remaining computationally efficient. Its cosmological evolution is computed using \texttt{CLASS} while its treatment of astrophysics is tied to the local star formation rate density, taking advantage of its exponential dependence on the density field to obtain the statistical properties of radiative fields.

The parametrization of \texttt{Zeus21} bears some similarities to that of \texttt{21cmFAST}. However, there are some important differences. One such difference lies in their parametrization of the fraction of galactic gas in stars $f_\star$. For \texttt{21cmFAST}, it is expressed as \cite{Park:2018ljd}
\begin{equation}
    f_\star^\texttt{21cmFAST}(M_{\rm h}) = 10^{f_{\star, 10}}\left(\frac{M_{\rm h}}{10^{10}M_\odot}\right)^{\alpha_\star},
    \label{eq:galactic_gas_21cmfast}
\end{equation}
where $M_{\rm h}$ is the halo mass, and the other parameters are defined in Table~\ref{tab:astro_params}. In other words, $f_\star^\texttt{21cmFAST}$ is modeled as a single power law of $M_{\rm h}$, whose slope is determined by $\alpha_\star$. On the other hand, \texttt{Zeus21} parametrizes $f_\star$ as the double power-law
\begin{equation}
    f_\star^\texttt{Zeus21}(M_{\rm h}) = \frac{2\epsilon_\star}{\left({M_{\rm h}}/{M_{\rm c}}\right)^{-\alpha_\star} + \left({M_{\rm h}}/{M_{\rm c}}\right)^{-\beta_\star}}, 
    \label{eq:galactic_gas_zeus21}
\end{equation}
where $\epsilon_\star$ is a normalization factor, $M_{\rm c}$ the critical mass of maximal fraction of galactic gas in stars, and $\beta_\star$ controls the high halo-mass power law. 

Confronted with a difference in parametrization, there are two different approaches that one can take. One is to simply run symbolic regression independently for each code, allowing differences in parametrization and variable names to propagate even if the underlying prescriptions are fundamentally similar. However, such an approach allows cosmetic differences to masquerade as mathematical discrepancies between symbolic regression results. This hinders our goal of comparing different codes on an equal footing later in the paper.
To overcome this, we take an alternative approach of bringing parametrizations into alignment ahead of time (analytically, prior to symbolic regression) where possible. Here, we modify the \texttt{Zeus21} power law by fixing the critical mass $M_{\rm c}$ to $10^{18}M_\odot$, four order of magnitudes larger than the maximal halo mass of $10^{14}M_\odot$ that the code simulates, and effectively reducing the problem to a single power law where the $\beta_\star$ term does not contribute (and was fixed to a fiducial value of $-0.5$). By setting Eqs.~\eqref{eq:galactic_gas_21cmfast} and \eqref{eq:galactic_gas_zeus21} equal and solving for $\epsilon_\star$, we get
\begin{equation}
    \epsilon_\star = \frac{10^{f_{\star, 10} + 8\alpha_\star}}{2}.
\end{equation}
Therefore, given $f_{\star, 10}$ and $\alpha_\star$ sampled from a Sobol sequence in the ranges prescribed by Table~\ref{tab:astro_params}, the corresponding $\epsilon_\star$ value can be used to run \texttt{Zeus21} while allowing us to express our results in the same parameters as those that \texttt{21cmFAST} uses. Of course, in pushing the \texttt{Zeus21} double power law to be effectively a single power law, we are in principle restricting the generality of \texttt{Zeus21}. However, in practice we find only a very weak (perhaps negligible) dependence of $\tau$ on $\beta_\star$.

For each point in parameter space, \texttt{Zeus21} outputs the hydrogen ionization history $\bar x_{\rm HII}(z)$ on a redshift grid spanning $5.5 \le z \le 30$ with spacing $\Delta z=0.05$. We generate 2048 simulations using Sobol-sequence sampling over the parameter ranges listed in Tables~\ref{tab:cosmo_params} and \ref{tab:astro_params}, using boxes of side length $300 \:{\rm Mpc}/h$ with a resolution of $1\,{\rm Mpc}/h$. Out of these, one run did not fully reionize by $z=5.5$, and three did not complete, and thus our dataset contains 2044 runs. All parameters not explicitly varied are fixed to their default \texttt{Zeus21} values. To compute $\tau$, we reconstruct the free-electron density from the simulated ionization history and numerically evaluate Eq.~\eqref{eq:tau}. We assume singly ionized helium traces hydrogen reionization, so that $x_e(z)=x_{\rm HII}(z)(1+x_{\rm He})$, where $x_{\rm He}\equiv n_{\rm He}/n_{\rm H}$ is the helium-to-hydrogen number density ratio, and model the second ionization of helium as an instantaneous transition at $z_{\rm HeIII}=3.5$, below which $x_e(z)=x_{\rm HII}(z)(1+2x_{\rm He})$. The electron density is then $\bar n_e(z)=x_e(z)\,n_{\rm H}(z)$, which we integrate over redshift to obtain $\tau$. 

Having run our simulations, we now have all relevant inputs that are needed to run a symbolic regression algorithm. At its barest essentials, this consists of a table of astrophysical and cosmological parameter values and their corresponding $\tau$ values predicted by a particular code. In order to guide symbolic regression towards physical solutions, in this table we substitute $\Omega_{\rm m}$ (normalized matter density) and $\Omega_{\rm b}$ (normalized baryon density) for $\Omega_{\rm m}h^2$ and $\Omega_{\rm b}h^2$, where $h$ is the Hubble parameter in units of $100\,\textrm{km}/\textrm{s}/\textrm{Mpc}$. This encourages the algorithm to output its formulae in terms of physical densities. We also enforce a common set of cosmological parameters, expressing the results from all three codes in terms $\Omega_{\rm m}h^2$ rather than $\Omega_{\rm c}h^2$. This also requires self-consistently reparameterizing \texttt{Zeus21} (using \texttt{CLASS}) in terms of $\sigma_8$ (root-mean-square fluctuations smoothed on $8h^{-1}\,\textrm{Mpc}$ scales) rather than its native $A_{\rm s}$ (primordial fluctuation amplitude). With a common set of cosmological parameters, we are able to more easily compare the results of symbolic regression fits for $\tau$.

\subsection{Symbolic Regression}
\label{subsec:meth:sr}

We aim to characterize the functional relationships between the optical depth $\tau$ and the cosmological $\Theta_\text{cosmo}$ and astrophysical $\Theta_\text{astro}$ parameters of interest described in Tables~\ref{tab:cosmo_params} and \ref{tab:astro_params}, respectively.
To this end, we employ symbolic regression (SR) to discover a closed-form mathematical expression $f$ such that $\tau = f(\Theta_\text{cosmo} \cup \Theta_\text{astro})$.

In this work, we utilize \texttt{PySR} \cite{Cranmer2023}, which is an open-source version of the \texttt{Eureqa} \cite{Eureqa} framework, representing the traditional approach to SR via evolutionary algorithms. In this framework, a population of expressions is randomly generated and evolved through selection based on fitness, crossover, and mutation to maintain diversity. Specifically, \texttt{PySR} utilizes ``islands" of isolated populations to prevent premature convergence. The algorithm evaluates performance using an $L_2$ loss after optimizing any free constants that may appear within the expressions. Output equations are given a complexity score, defined as the number of nodes in an expression tree encoding an equation \cite{Cranmer2023}. In our context, it serves as a proxy for equation length and can be used to strike a balance between equation simplicity and accuracy.

Our procedure involves running \texttt{PySR} with default hyperparameters, and a maximum complexity of 90. The library of elementary functions (or function set) comprises $\{ +, -, \times, /, \placeholder^\placeholder, \exp, \log \}$. The complexity of the exponent in the power operator is limited to 10, as we found this facilitates the discovery of more accurate and interpretable expressions. Our data is partitioned using an 80/20 train--test split in order to evaluate possible overfitting \textit{a posteriori}. Results are presented in the form of a Pareto front: in our context, a curve of complexity versus the mean squared error over parameter space between our fits and the ground truth data.

Supplementing the automatic exploration of \texttt{PySR} is our knowledge of reionization. The optical depth $\tau$ is directly proportional to $\Omega_{\rm b}h^2$ (excepting $\Omega_{\rm b}h^2$'s minor impact on astrophysical processes) because $n_e$ is proportional to $\Omega_{\rm b}h^2$ (see Eq.~\ref{eq:tau}). While SR is able to find this dependence on its own in most runs, we found that fitting to $\tau/\Omega_{\rm b}h^2$ for \texttt{21cmFAST} and \texttt{Zeus21} yielded more accurate expressions because SR will then only generate and iterate over expressions obeying this dependence, instead of also having to produce and potentially later reject invalid expressions. We stress that this does not restrict \texttt{PySR}'s freedom to find higher-order dependencies on $\Omega_{\rm b}h^2$, and we find that this does indeed happen for \texttt{21cmFAST}. \\

\section{Symbolic regression expressions}\label{sec:results}

\subsection{Accurate fits to the optical depth}

\begin{figure}
    \centering
    \includegraphics[width=\linewidth]{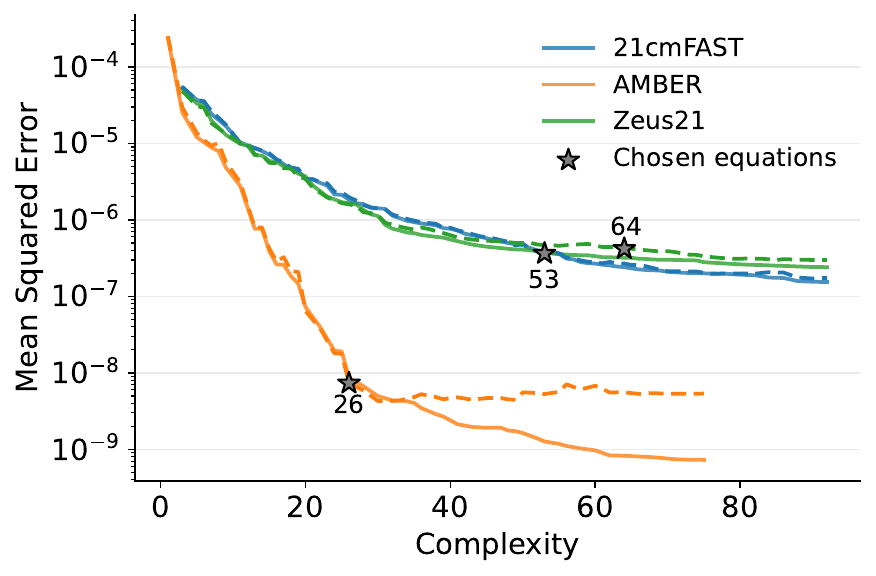}
    \caption{Pareto front obtained with \texttt{PySR} for $\tau$ for the three simulations considered. Solid lines represent training set performance, and dashed lines corresponding to the test sets. \texttt{21cmFAST} and \texttt{Zeus21} equations encode meaningful physics even at long complexity, while \texttt{AMBER} behavior is best captured by short equations, as seen by the overfitting for complexities above 30. The selected \texttt{21cmFAST} and \texttt{Zeus21} equations have respective complexities of 53 and 64 (which we adjusted to represent the complexity on $\tau$ rather than on $\tau/\Omega_{\rm b}h^2$), while the \texttt{AMBER} equation has complexity 26.}
    \label{fig:pareto_tau}
\end{figure}

Our runs show that the physics controlling $\tau$ can be captured by compact equations across all cosmological simulations. The Pareto front for equations regressed from the cosmological simulations is presented in Fig.~\ref{fig:pareto_tau}. Solid curves show the results from training runs while dashed curves are from test runs, providing a visual check on any possible overfitting. Stars indicate the symbolic regression expressions that are selected for further use later in the paper.

Each code behaves in a different way when fit by symbolic regression. \texttt{21cmFAST} is best described by large equations, as shown by the lack of significant overfitting. However, the mean squared error decreases at a slower rate once the complexity of the $\tau$ expression reaches 53; we therefore select this version of the $\tau$ fit as it best balances length with accuracy. It is given by
\begin{widetext}
\begin{equation}
\begin{aligned}
\tau_\texttt{21cmFAST} ={} \Omega_{\rm b}h^2
& \left[ \left(
\frac{M_{\rm turn} - a_1}
{\Omega_{\rm b}h^2 + a_2\,e^{a_3(\alpha_{\rm esc} + \alpha_{\star})}}
- a_4
\right) \right. 
\left. \left(
- a_5(\alpha_{\rm esc}
+\alpha_{\star}+
f_{\rm esc,10})
+ \frac{\Omega_{\rm m}h^2 + n_{\rm s}}
{a_6 \,f_{\rm esc,10} + f_{\star,10} - a_7}
\right) \right]^{n_{s}} \\
& \times \left[
\sigma_{8}
+ h\,\log\left(
-\frac{1}{\left(R_{\rm bubble}^{\rm max}\right)^{a_8}}
- \Omega_{\rm b}h^2
+ a_9
\right)
\right],
\label{eq:tau_21cmf}
\end{aligned}
\end{equation}
where $\tau_\texttt{21cmFAST}$ is the modeled \texttt{21cmFAST} optical depth, and the constants are $\mathbf{a}$ = [9.7452345, 0.18550034, 2.6163442, 9.55555, 0.12663826, 0.50611649, 0.46820813, 1.3906883, 0.6873837]. Note that we used \texttt{sympy} \cite{sympy} to simplify \texttt{PySR}'s raw output expressions.

Equations obtained for \texttt{Zeus21} overfit slightly at large complexities, but error on the test dataset still decreases with training error, meaning meaningful physics can still be extracted from the longer equations. Similarly to \texttt{21cmFAST}, a plateau in error is reached at high complexity. The parameter $\Omega_{\rm m}h^2$ appears for the first time at a complexity of 49, while all other parameters are used in shorter equations. This reveals that it is the least important parameter in determining $\tau$ to the precision of the SR-generated equations, as it is only employed in large equations which are only marginally more accurate than the shorter ones.  We choose to study the equation of complexity 64 as it is the shortest equation producing reliable derivatives for $\Omega_{\rm m}h^2$, which is important for parameter forecasts and to trust the physics our equations describe (see Sections~\ref{subsec:res:derivatives} and \ref{sec:fisher_forecast}). The selected equation is
\begin{equation}
\begin{aligned}
\tau_{\texttt{Zeus21}} ={}&
\Omega_{\rm b}h^2\Biggl[-h+b_1\biggl(
b_2(\alpha_{\rm esc}+\alpha_{\star}-h)
+b_2\Bigl[(\Omega_{\rm m}h^2+n_{\rm s})
\Bigl[(\alpha_{\rm esc}+\alpha_{\star}-b_3)\,
\bigl(M_{\rm turn}+\alpha_{\rm esc}+\alpha_{\star}-b_7
\\
&-(\alpha_{\rm esc}+\alpha_{\star}-b_6)\,
\bigl[f_{\rm esc,10}+(M_{\rm turn}-b_4)^{b_5}\bigr]\bigr)
+b_8\Bigr]^{\frac{f_{\rm esc,10}+f_{\star,10}+b_9}
{\log(M_{\rm turn})}}
\Bigr]^{\sigma_8 n_{\rm s}}
\biggr)^{b_{10}}-b_{11}\Biggr],
\end{aligned}
\label{eq:tau_zeus21}
\end{equation}
where $\tau_{\texttt{Zeus21}}$ is the modeled
\texttt{Zeus21} optical depth, and the constants are
$\mathbf{b}$ = [3.306756, 0.41762363, 0.6167628,
7.9879537, 1.6699086, 0.3520325, 10.084869, 3.815524,
4.8786626, 0.56289685, 1.3836839]. 
\end{widetext}

Unlike the first two simulations, \texttt{AMBER} equations show a rapid decrease in MSE until a complexity of 30, after which adding extra terms to the equation systematically over-fits the data. This shows \texttt{AMBER} is the most lightweight code of the three to parametrize as short equations capture all the essential physics. We choose the equation of complexity 26 as its structure maximizes interpretability, while remaining accurate enough for our study. Our \texttt{AMBER} equation is given by

\begin{equation}
\scalebox{0.92}{$
\displaystyle
    \tau_\texttt{AMBER} = \frac{\Omega_{\rm b}h^2\left(c_1 z_\text{mid} + c_2\right)\left(z_\text{mid} + c_3\left(\Delta_z\ln{\left(A_z \right)}^{c_4}\right)^{c_5} - c_6\right)}{\Omega_{\rm m}h^2 + c_7},
$}
\label{eq:tau_amber}
\end{equation}

\noindent where $\tau_\texttt{AMBER}$ is the modeled \texttt{AMBER} optical depth, and the constants are $\mathbf{c}$ = [0.003642231, 0.08058576, 0.10047147, 0.67381823, 1.1245414, 0.932408, 0.15851374]. The Pareto front obtained with \texttt{PySR} therefore showcases the ability of compact equations to predict $\tau$ across several cosmological simulations, and also reveals that the astrophysical parameters employed in \texttt{AMBER} allow for a much simpler parametrization of $\tau$ than in the other two codes.

\begin{figure}
    \centering
    \includegraphics[width=\linewidth]{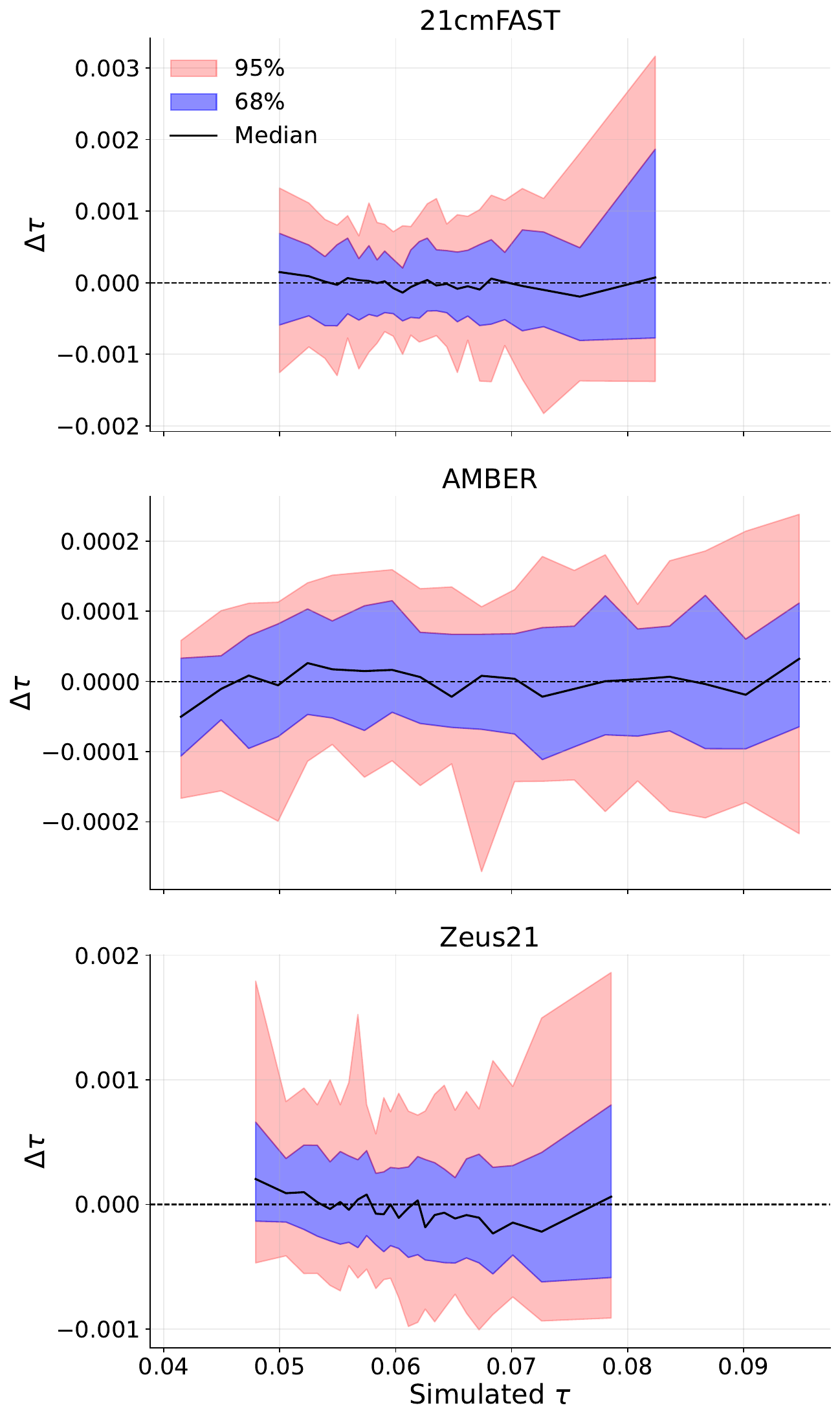}
    \caption{Residuals on all data (training and test) of \texttt{PySR}'s equations for all simulations. The blue and red regions are the 16-to-84 (68\%) and the 2.5-to-97.5 (95\%) percentile zones, respectively. \texttt{21cmFAST}, \texttt{AMBER}, and \texttt{Zeus21} have RMSEs of $6.1\times10^{-4}$, $9.1\times10^{-5}$, and $5.8\times10^{-4}$ respectively on all data (training and test combined). For both \texttt{21cmFAST} and \texttt{AMBER}, errors are symmetric which shows that our functions do not systematically overestimate or underestimate $\tau$. Our \texttt{Zeus21} fit systematically overestimates $\tau$ when it is lower than 0.055, and underestimates it for $0.065 \leq \tau \leq 0.075$.}
    \label{fig:residuals_tau}
\end{figure}

In general, our symbolic regression fits are extremely accurate, surpassing current observational limitations. We plot the residuals of our chosen equations in Fig.~\ref{fig:residuals_tau}. Across all simulated $\tau$ values, the error on the \texttt{21cmFAST} and \texttt{Zeus21} fits is $O(10^{-3})$, and the one on \texttt{AMBER} is $O(10^{-4})$. As current observational uncertainties on $\tau$
are of order $10^{-2}$ \cite{Tristram:2023haj},
the fitting errors are considerably smaller than
the present observational uncertainty. The greatest error on $\tau$ for \texttt{21cmFAST} is in the $\tau \geq 0.075$ bin, which is roughly twice as large as the errors observed in the rest of $\tau$-space. As all plotted bins have an equal amount of data, the large bin width indicates a lack of training data in the high-$\tau$ region. This causes SR to prioritize fitting to the bulk of the data at lower $\tau$, explaining the higher error. A similar phenomenon is observed in the \texttt{Zeus21} data, with errors growing at the extrema in $\tau$. Despite these increases, accuracy remains well below the current observational limits and the fits remain reliable across the entire parameter space. SR is therefore able to accurately link $\tau$ to our simulations' input parameters.

The expressions generated with symbolic regression are also partly interpretable and encode meaningful physics. One place where this becomes apparent is in the identification of degeneracies between input parameters. In particular, in the expressions for $\tau_\texttt{21cmFAST}$ (Eq.~\ref{eq:tau_21cmf}) and $\tau_\texttt{Zeus21}$ (Eq.~\ref{eq:tau_zeus21}), $\alpha_\star$ and $\alpha_{\rm esc}$ are fully degenerate as they are always summed with one another. We also see a partial degeneracy between $f_{\star, 10}$ and $f_{\rm esc,10}$. These can be explained using the definitions of these pairs of parameters: $f_{\star, 10}$ and $\alpha_\star$ jointly determine the fraction of galactic gas in stars $f_\star$ through Eq.~\eqref{eq:galactic_gas_21cmfast}, while $f_{\rm esc, 10}$ and $\alpha_{\rm esc}$ determine the escape fraction of UV photons from galaxies $f_{\rm esc}$ \cite{Park:2018ljd}. Therefore, the aforementioned parameter degeneracies are simply a manifestation of the fact that star formation and escape fraction can be traded off one another to keep the ionizing photon budget (and therefore the ionization history) the same. We thus see that key physics can be recovered from examining parameter degeneracies present in our symbolic regression-produced equations.

Understanding the symbolic regression results from \texttt{AMBER} require a different strategy, since \texttt{AMBER}'s parameters are more phenomenological. Still, we see key features being captured by Eq.~\eqref{eq:tau_amber}. For example, one sees that $z_{\rm mid}$ is the main driver in determining $\tau$, as expected (since $\tau$ measures an integrated history). The asymmetry and duration provide smaller corrections, with the asymmetry parameter $A_z$ only affecting the results if $\Delta_z > 0$. Essentially, the asymmetry in ionization history can only have an effect if the ionization history possesses long tails on at least one side of $z_{\rm mid}$, and not if reionization is instantaneous. These results showcase the way in which the “transparency" of analytic equations (compared to black-box machine learning methods) can be an attractive motivator for using symbolic regression to summarize complex cosmological simulations.

Of course, the relative merits of black-box machine learning methods and symbolic regression cannot be judged on interpretability alone; accuracy must also be considered. Fortunately, we find that the equations produced by \texttt{PySR} show competitive accuracy compared to neural networks trained on the same data. To establish this, we trained a \texttt{PyTorch} \cite{Paszke2019} neural network (NN) consisting of 3 layers with 64 neurons using \texttt{SiLU} activation functions. Inputs were standardized using the mean and variance of training datasets, and training was performed over 20,000 epochs with a batch size of 256 using the \texttt{AdamW} optimizer, with a learning rate of $3\times10^{-4}$, a weight decay of $1\times10^{-4}$, and a 10\% validation split for model selection. The mean-squared errors on the test datasets are presented in Table~\ref{tab:nn_errors}. The NN's errors on the \texttt{Zeus21} data are similar to the ones obtained with SR. On \texttt{AMBER} data, SR's MSE is a factor of three lower than the one obtained with the NN, while on the \texttt{21cmFAST} data, the NN outperforms SR by a factor of five. It is possible that the optical depth's functional dependencies for \texttt{21cmFAST} are too complex to be \textit{fully} discovered by SR, or that \texttt{PySR} is simply unable to fully resolve it, meaning there is room for future improvement in such equation-discovery algorithms. On the whole, the errors in the SR-produced functions are similar to those obtained with neural networks, even outperforming them on \texttt{AMBER} data.

\begin{table}[!h]
\setlength{\tabcolsep}{6pt}
\centering
\caption{Mean-squared error (MSE) of the neural network on the test datasets, compared to the errors of symbolic regression. The errors of the SR and the NN predictions are of the same order of magnitude for a given simulation.}
\label{tab:nn_errors}
\renewcommand{\arraystretch}{1.15}
\begin{tabular}{l l l}
\toprule
Simulation & SR MSE & NN MSE \\
\midrule
\texttt{21cmFAST} & $3.6\times10^{-7}$ & $7.5\times10^{-8}$ \\
\texttt{AMBER} & $7.3\times10^{-9}$ & $2.4\times10^{-8}$ \\
\texttt{Zeus21} & $4.2\times10^{-7}$ & $2.9\times10^{-7}$ \\
\bottomrule
\end{tabular}
\end{table}

\subsection{Accurate first derivatives}\label{subsec:res:derivatives}

On top of accurately predicting $\tau$ itself, it is important that our expressions produce accurate first order partial derivatives with respect to parameters. An example application of this would be forecasts using Fisher matrices, which depend crucially on these derivatives. Beyond parameter forecasts, differentiable codes are extremely convenient in any physics problem which can be recast as finding the extrema of a function. SR functions which properly represent the derivatives of cosmological simulation outputs make this quantity fully differentiable, which is not the case when solely utilizing the simulation.

In order to evaluate the match between analytic derivatives (obtained by differentiating SR expressions) and numerical derivatives (obtained directly from simulations), we select five different points in parameter space on which to evaluate them. By selecting a reasonably broad range of $\tau$ values, we strike a balance between an exhaustive test of the numerical derivatives and possible accidents of good (or bad) agreement. From our \texttt{21cmFAST} dataset, we identified parameter chains producing a fiducial $\tau$ value of 0.054 \cite{Planck:2018vyg}, a low $\tau = 0.050$, a high $\tau = 0.090$, and two randomly selected points that have $\tau = 0.062$ and 0.074. We obtain the output AMBER astrophysical parameters $z_{\rm mid}, \Delta_z, \textrm{and } A_z$ from those exact five 21cmFAST runs (as they can be directly computed from the output ionization curve) and evaluate AMBER derivatives on the corresponding parameter chains. \texttt{Zeus21} fiducial points are selected directly from our \texttt{Zeus21} dataset, since the difference in cosmological parameter prescriptions between the codes (some using $\Omega$ and others $\Omega h^2$ as inputs) means points from the \texttt{21cmFAST} dataset are not guaranteed to lie within the \texttt{Zeus21} training range when transforming $\Omega_{\rm m}$ and $\Omega_{\rm b}$ into densities. We therefore select the points in the \texttt{Zeus21} dataset producing optical depths of 0.054, 0.050, and 0.090, as well as two randomly selected points with $\tau = 0.066$ and 0.054 (with different parameters than the first $\tau=0.054$ fiducial).

\begin{figure*}
    \centering
    \includegraphics[width=\linewidth]{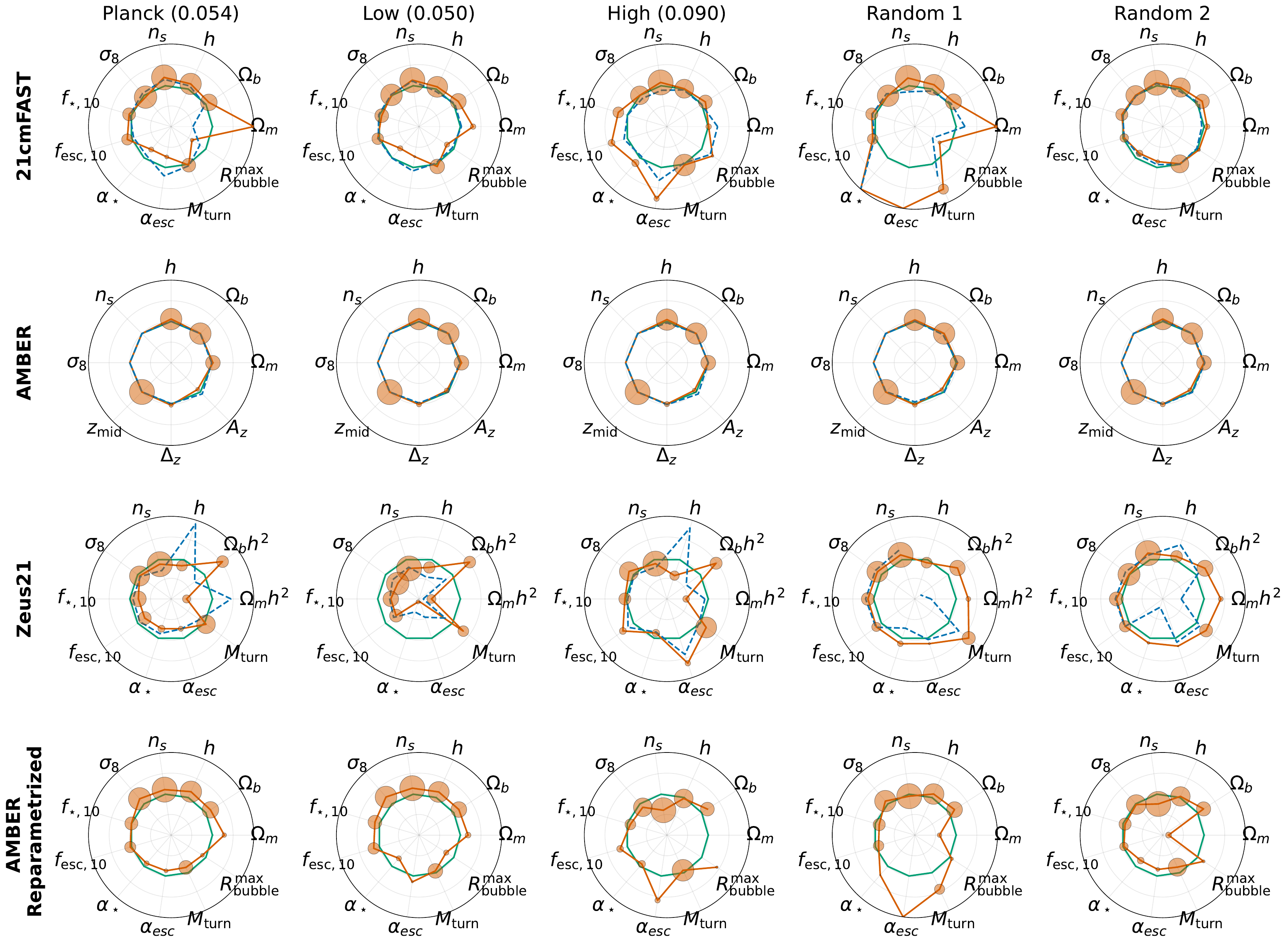}
    \caption{Plot of the ratio between analytic and numerical derivative values (in orange) on five fiducial points. On this curve is overlayed a circle whose size is proportional to $|\theta\times\partial_\theta\tau|$, so that a higher size indicates a more impactful parameter in determining $\tau$. The green line is the unit circle and serves as reference for where a perfect agreement between analytic and numerical derivatives lies. In dashed-blue is the ratio between the neural network derivatives and the numerical derivatives. Missing data points indicate a negative ratio (i.e. a misestimated derivative sign). The bottom panel corresponds to \texttt{AMBER} derivatives computed through Eq.~\eqref{eq:chain-rule}, where we have reparametrized the \texttt{AMBER} expression as a function of \texttt{21cmFAST} variables. Ratios exceeding 2 only occur in \texttt{21cmFAST} and reparametrized-\texttt{AMBER} derivatives and are clipped in order to better represent details near the unit circles, but no data exceeds a ratio of 2.90.}
    \label{fig:sr-derivatives}
\end{figure*}

Fig.~\ref{fig:sr-derivatives} summarizes the performance of our symbolic derivatives in predicting numerical derivatives. Marker sizes indicate the importance/magnitude of different parameter derivatives (as quantified in a dimensionless way by the derivative with respect to the logarithm of each parameter). The green line shows perfect agreement, and the periphery indicates a factor-of-two (or greater) disagreement. Broadly speaking we see that our analytic derivatives are generally trustworthy. Our fits to \texttt{AMBER} perform near-perfectly on predicting first-order derivatives no matter the parameter importance in determining $\tau$. \texttt{21cmFAST} also produces accurate derivatives with respect to the most important parameters in determining $\tau$ but can struggle to predict lower-amplitude derivatives. In fact, the ratio between analytic and numerical derivatives for parameters such as $\sigma_8$, $n_s$, $h$ and $\Omega_{\rm b}$ are close to 1 across all fiducial points, but parameters like $\alpha_\star$ and $\alpha_\text{esc}$ are misestimated at many points. This is most noticeable at the point with $\tau=0.062$ where our equations' derivatives with respect to these two parameters are about 2.5 times larger than the true simulated value. Similar conclusions can be drawn for the \texttt{Zeus21} derivatives: $n_{\rm s}$, $\sigma_8$, $f_{\star, 10}$, $f_{\rm esc, 10}$, and $M_{\rm turn}$ have accurate derivatives at most fiducial points, while $\alpha_\star$, $\alpha_\text{esc}$, and $\Omega_{\rm m}h^2$ derivatives are often misestimated. What is different between our \texttt{Zeus21} and \texttt{21cmFAST} fits are that the former has $\Omega_{\rm b}h^2$ derivatives are systematically overestimated, and the low-$\tau$ (0.050) derivatives seem systematically underestimated across most parameters. On the \texttt{21cmFAST} and \texttt{Zeus21} datasets, SR has trouble in fully resolving the behaviour of parameters which only impact $\tau$ weakly, but we find that it is still able to get the correct sign and order of magnitude of the derivatives. Our function can still reliably be used in many applications requiring derivatives such as Fisher forecasts, where errors will be suppressed by low derivative magnitudes. For both these simulations, SR is therefore able to produce reliable first-order derivatives for the parameters which impact $\tau$ the most, and gives strong estimates for the sub-dominant ones.

The comparable performance between our expressions and neural networks remains true for first-order derivatives. For \texttt{21cmFAST}, NN derivatives are as accurate as SR's for the parameters of most importance, and outperform SR when predicting low-amplitude derivatives (see Fig. \ref{fig:sr-derivatives}). Our neural networks are therefore preferred when predicting overall \texttt{21cmFAST} derivative accuracy, but the similar accuracy on the most constraining parameters means Fisher forecasts should not be affected by the method used to produce derivatives. As for \texttt{AMBER}'s derivatives, the NN is extremely accurate, matching SR performance. \texttt{Zeus21} SR and NN derivatives are of similar accuracy, and even both systematically underpredict derivative magnitudes at the $\tau=0.050$ fiducial. Our SR expressions are therefore competitive alternatives to simple neural networks for predicting first derivative accuracy, while retaining far more transparency in that they can be computed by differentiating an analytic expression for $\tau$.

We have shown that symbolic regression produces accurate equations for $\tau$ and its first order derivatives which can compete with machine learning algorithms. We will now show that SR also enables a natural comparison between these different simulations through careful analysis of the equations it produces for $\tau$.

\section{Using symbolic regression to compare cosmological simulations}\label{sec:simulation_comparison}

The complexity of the SR expressions provides a useful
measure of how compactly the optical depth dependence
of each simulation can be represented analytically.
Out of the three codes, \texttt{AMBER} is the “simplest" one in terms of modeling reionization, in the sense that its optical depth $\tau$ can be modeled the most accurately and using the simplest equations. In fact, Fig. \ref{fig:residuals_tau} shows that \texttt{AMBER}'s RMSE is roughly 10 times lower than \texttt{21cmFAST}'s and \texttt{Zeus21}'s. What is even more informative is the fact MSE on the test set reaches a minimum at a complexity of 30 (see Fig. \ref{fig:pareto_tau}), after which longer, more complex equations overfit to data and do not encode any more physics of \texttt{AMBER}. This supports the fact that \texttt{AMBER} encodes reionization with parameters which can be arranged in a simple equation to encode almost all of its physics that governs $\tau$. This is not true of \texttt{21cmFAST} and \texttt{Zeus21} as adding complexity reduces MSE without overfitting to data, meaning $\tau$'s analytic dependencies are inherently complex and may not be entirely discoverable in the form of a compact analytic equation. We speculate that this is due to the input parameters native to these codes. \texttt{AMBER}'s parameters $z_\text{mid}, \Delta_z$, and $A_z$ determine entirely the ionization history, which in turn entirely determines $\tau$ through Eq. \ref{eq:tau}. Should the ionization history be a simple function of these parameters, then $\tau$ could be parametrized using only a few parameter combinations. On the other hand, the astrophysical parameters employed by \texttt{21cmFAST} and \texttt{Zeus21} may only be related to the ionization history with complex dependencies, meaning $\tau$ would be more difficultly expressed as a simple combination of such parameters. \texttt{AMBER} is therefore the most “simple" code in terms of the relationship between $\tau$ and its input parameters, but this is more a statement of its choice of input parameters rather than a statement of simulation simplicity or accuracy (compared to reality).

Beyond comparing equation error and complexity between codes, symbolic regression provides a framework to directly compare simulations' dependencies on a common set of input variables. The steps described in Section~\ref{subsec:meth:sims} have already ensured that \texttt{21cmFAST}, \texttt{AMBER}, and \texttt{Zeus21} share cosmological parameters and that \texttt{21cmFAST} and \texttt{Zeus21} have a harmonized set of astrophysical parameters. What remains is to relate the \texttt{AMBER} parameters $z_\text{mid}$, $\Delta_z$ and $A_z$ to these parameters. We do so by computing $z_\text{mid}$, $\Delta_z$ and $A_z$ in post-processing from a \texttt{21cmFAST} output ionization history; there is therefore a correspondence between a given set of inputs $\boldsymbol\theta_\texttt{21cmFAST}$ and \texttt{AMBER} inputs $[z_\text{mid}, \Delta_z, A_z]$. We gather these quantities from all of our \texttt{21cmFAST} parameter chains, and use \texttt{PySR} to find functions of the form $z_\text{mid} = f(\boldsymbol{\theta_\texttt{21cmFAST}})$ (similarly for $\Delta_z$ and $A_z$) allowing us to then re-express $\tau_\texttt{AMBER}$ (Eq. \ref{eq:tau_amber}) as a function of \texttt{21cmFAST} input variables. We run \texttt{PySR} limiting $z_\text{mid}$ to a complexity of 90 and $\Delta_z, A_z$ to 50. The selected equation for $z_\text{mid}$ has complexity 51 and is
\begin{align}
z_{\rm mid} =\;&
\left(d_1 - \Omega_{\rm b}h^2\right)
\left(- \alpha_{\rm esc} + \alpha_{\star}
+ e^{\frac{\Omega_{\rm m}h^2}{h} + \sigma_{8} + n_{\rm s}}\right)
\nonumber \\
&+ \left(d_2 - d_3\,\alpha_{\rm esc}\right)
\left(\Omega_{\rm m}h^2 + f_{\rm esc,10} + f_{\star,10}\right)
\nonumber \\
&- \left(d_4 - M_{\rm turn}\right)
\log\!\left(\alpha_{\star}^{{d_5}^{\alpha_{\rm esc}}} + d_6\right)
- \frac{d_7}{R_{\rm bubble}^{\rm max}},
\label{eq:z_mid}
\end{align}
\noindent where the constants are $\mathbf{d}$ = [1.7278208, 3.0577748, 1.67564561368761, 9.715671, 0.076510504, 0.057638414, 5.591118]. The selected equations for $\Delta_z$ and $A_z$ have complexities of 25 and 26, respectively, and are
\begin{equation}
\begin{aligned}
\Delta_z ={}&
\sigma_{8}
\Biggl[
\left(
-\alpha_{\rm esc}
-\alpha_{\star}
+k_1
\right)
\left(
-M_{\rm turn}
-f_{\rm esc,10}
-f_{\star,10}
+k_2
\right)
\\
&\qquad
+k_3
+\frac{k_4}{R_{\rm bubble}^{\rm max}}
\Biggr]^{\Omega_{\rm m}h^2+n_{\rm s}}
\label{eq:z_dur}
\end{aligned}
\end{equation}
for the duration of reionization, with $\mathbf{k}$ = [1.0441043, 8.513224, 1.9412702, 14.433918], and 
\begin{equation}
    A_z = \exp\left[{\frac{{m_1}^{n_{\rm s}} + R_{\rm bubble}^{\rm max}\Omega_{\rm m}h^2}{M_{\rm turn} + f_{\rm esc,10} + f_{\star,10} + \left(\alpha_{\rm esc} + \alpha_{\star} + m_2\right)^{m_3}}}\right] - m_4
\label{eq:z_asy}
\end{equation}
for its asymmetry, with $\mathbf{m}$ = [5.7143917, 0.52121097, 2.8415148, 0.38884723]. 

It is interesting to note that similar parameter degeneracies between $\alpha_\star$ and $\alpha_{\rm esc}$, as well as between $f_{\star, 10}$ and $f_{\rm esc, 10}$ are observed in the above expressions. These expressions are trained on \texttt{21cmFAST} data, which contains these degeneracies in its expression for $\tau$ (see Eq. \ref{eq:tau_21cmf}). It is possible that the degeneracies on $\tau$ stem from the degeneracies on $z_{\rm mid}, \Delta_z$, and $A_z$, which are then represented in the simulation's output $\tau$.

The derivatives of $\tau_\texttt{AMBER}$ with respect to \texttt{21cmFAST} inputs are as accurate as those of $\tau_\texttt{21cmFAST}$, despite the added error due to reparameterization. Evaluating such derivatives is done using the chain rule:

\begin{equation}
    \begin{aligned}
& \left.\frac{\partial\tau}{\partial\theta_i}\right|_{\boldsymbol{\theta}_{\texttt{21cmFAST}}}
=
\left.\frac{\partial\tau}{\partial\theta_i}\right|_{\boldsymbol{\theta}_{\texttt{AMBER}}}
+ \left.\frac{\partial \tau}{\partial z_\text{mid}}\right|_{\boldsymbol{\theta}_{\texttt{AMBER}}}
  \left.\frac{\partial z_\text{mid}}{\partial\theta_i}\right|_{\boldsymbol{\theta}_{\texttt{21cmFAST}}} \\
\quad&
+ \left.\frac{\partial \tau}{\partial \Delta_z}\right|_{\boldsymbol{\theta}_{\texttt{AMBER}}}
  \left.\frac{\partial \Delta_z}{\partial\theta_i}\right|_{\boldsymbol{\theta}_{\texttt{21cmFAST}}}
+ \left.\frac{\partial \tau}{\partial A_z}\right|_{\boldsymbol{\theta}_{\texttt{AMBER}}}
  \left.\frac{\partial A_z}{\partial\theta_i}\right|_{\boldsymbol{\theta}_{\texttt{21cmFAST}}}
\end{aligned}
\label{eq:chain-rule}
\end{equation}
where it is understood that $\boldsymbol{\theta_\texttt{21cmFAST}}$ and $\boldsymbol{\theta_\texttt{AMBER}}$ represent the same point in parameter space, expressed in their native variable set. It is preferable to use Eq.~\eqref{eq:chain-rule} rather than direct differentiation because numerical derivatives of \texttt{AMBER} with respect to $\boldsymbol{\theta_\texttt{21cmFAST}}$ can also be computed in this way, which cannot be directly done from the simulation. Our \texttt{AMBER}-trained expression's derivatives with respect to \texttt{21cmFAST} input parameters are presented in the bottom panel of Fig.~\ref{fig:sr-derivatives}. The derivative accuracy decreases heavily compared to the one on \texttt{AMBER}'s native variables because of the errors induced in the reparametrization. However, Fig. \ref{fig:sr-derivatives} shows the accuracy is comparable to the \texttt{21cmFAST} expression's derivatives, with a similar trend of performing better on the parameters which constrain $\tau$ the most. The \texttt{AMBER} derivatives with respect to \texttt{21cmFAST} input parameters are therefore reliable enough to be trusted when comparing derivative magnitudes between $\tau_\texttt{AMBER}$ and our other expressions, when all are expressed in the same variable set.

\begin{figure*}
    \centering
    \includegraphics[width=\linewidth]{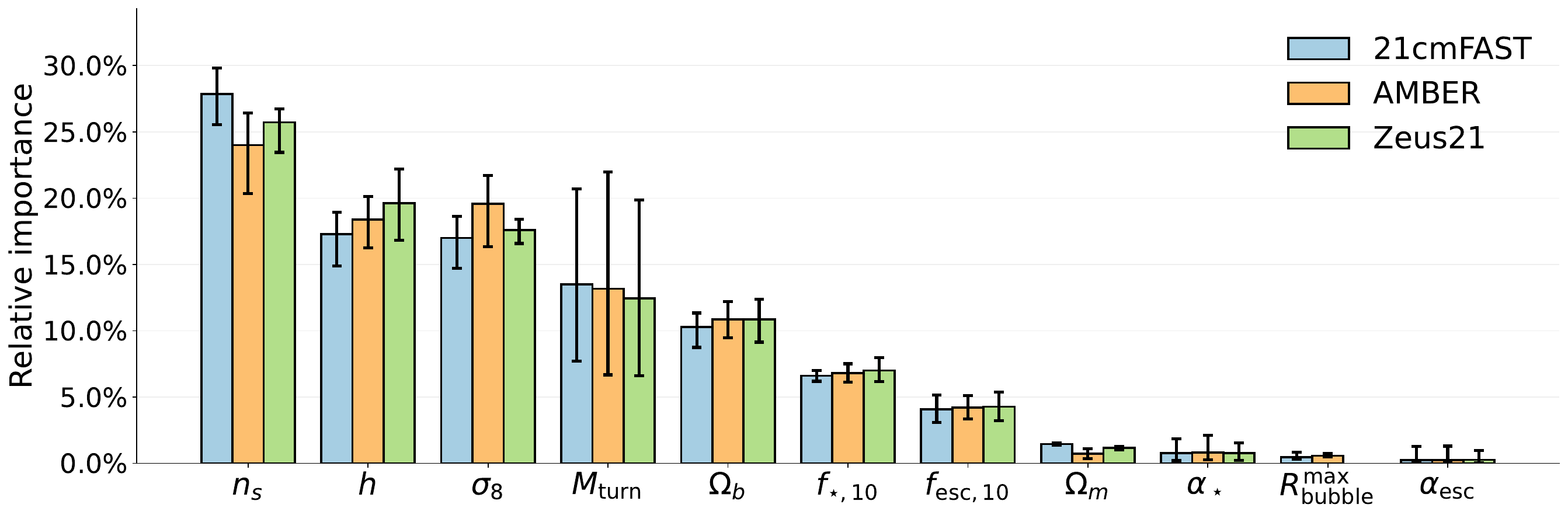}
    \caption{Mean relative importance of astrophysical and cosmological parameters on the \texttt{21cmFAST}-, \texttt{AMBER}-, and \texttt{Zeus21}-trained $\tau$ expressions, computed over combined training and test datasets. The black bars span the 16th to 84th percentile of values observed across parameter space. All three expressions show high agreement as to the most important parameters in determining $\tau$, as measured by normalized derivative magnitudes.}
    \label{fig:relative_importance}
\end{figure*}

Having reexpressed $\tau_\texttt{AMBER}$ as a function of \texttt{21cmFAST} input variables, our three cosmological simulations can be efficiently compared as they are all expressed in a common basis (with the caveat that $\tau_\texttt{21cmFAST}$ and $\tau_\texttt{AMBER}$ use $R_{\rm bubble}^{\rm max}$, which is not a native parameter in \texttt{Zeus21}). While directly comparing functional forms beyond degeneracies is a difficult task given the complexity of the equations, derivative magnitudes hold information about the relative importance of parameters in determining $\tau$. We normalize partial derivative magnitudes by the parameter $\theta_i$'s value, yielding the relative importance $|\theta_i\times\partial_{\theta_i}\tau|$ that was previously used to visually represent parameter importance in Figure~\ref{fig:sr-derivatives}. We compute the relative importance of each parameter across the 2025 parameter chains of the \texttt{21cmFAST} dataset for $\tau_\texttt{21cmFAST}$ and $\tau_\texttt{AMBER}$, and across the 2044 \texttt{Zeus21} realizations for $\tau_\texttt{Zeus21}$, normalizing values at each fiducial point so that they sum to a total importance of 100\%. These results are presented in Figure~\ref{fig:relative_importance}.

Overall, these results show that the three simulations
give broadly consistent rankings of the parameters
most important for determining $\tau$, despite
differences in their numerical implementations and
parameterizations. The most important cosmological parameters are $n_{\rm s}$, $h$, $\sigma_8$, and $\Omega_{\rm b}$, and the most important astrophysical parameters are $M_{\rm turn}$, $f_{\star, 10}$, and $f_{\rm esc, 10}$. The remaining parameters, $\Omega_{\rm m}$, $\alpha_\star$, $\alpha_{\rm esc}$, and $R_{\rm bubble}^{\rm max}$ are more peripheral parameters that do not strongly affect $\tau$ for any of our three models. On the whole, these results support the claim that \texttt{21cmFAST}, \texttt{Zeus21}, and \texttt{AMBER} agree on which parameters determine $\tau$ and hence the overall ionization history of the universe, despite their differences in numerical implementations and parameterizations.

Furthermore, as our expression for $\tau_\texttt{AMBER}$ contains symbolic-regression expressions trained on \texttt{21cmFAST} data (i.e. the expressions for $z_{\rm mid}, \Delta_z$, and $A_z$), agreement between the simulations shows that they both agree on how the ionization history impacts $\tau$ values. When evaluating both expressions on the \texttt{21cmFAST} dataset, the RMS difference in answers between the two formulas is $\sim 6.1\times10^{-4}$, which corresponds to the RMSE of $\tau_\texttt{21cmFAST}$. This shows that $\tau_\texttt{AMBER}$ can accurately predict \texttt{21cmFAST} outputs up to the precision of the \texttt{21cmFAST}-trained SR function, when reparametrizing the ionization history parameters as a function of \texttt{21cmFAST} variables. The expression for \texttt{AMBER} can therefore be seen as a more compact way to store all of the information required to compute $\tau$ for \texttt{21cmFAST}, even though $\tau_\texttt{AMBER}$ was trained on \texttt{AMBER}'s ionization history rather than on \texttt{21cmFAST}'s.

\section{Cosmological forecasts from the optical depth relations}
\label{sec:fisher_forecast}

As an application of the SR relations, we examine how a model for $\tau$ changes cosmological parameter forecasts. Except on the smallest and largest scales,  the CMB mostly easily constrains the combination $A_{\rm s}e^{-2\tau}$, so the uncertainty in $\tau$ limits the inferred primordial amplitude $A_s$, even though some degeneracy breaking is provided by large-scale polarization or CMB lensing measurements. Ref.~\cite{Liu:2015txa} showed that this degeneracy can be further broken using a self-consistent reionization model that predicts $\tau$ from other parameters. Our closed-form relations for $\tau$ make it relatively simple to extend this treatment in two ways. First, the availability of analytic expressions facilitates broad explorations of parameter space, moving away from typical Fisher studies where forecasts are performed about just a single point in parameter space. Second, with our expressions from Section~\ref{sec:simulation_comparison} for placing different reionization models on a common parametrization, one can perform controlled tests of the model dependence of forecasts.

\subsection{Forecast setup}
\label{subsec:fisher_methods}
 
We consider two fiducial models with optical depths
$\tau\simeq0.054$ and $0.090$. The former is motivated
by the Planck 2018 constraint on $\tau$
\cite{Planck:2018vyg}, while the latter is representative of discussions revolving around a higher optical depth that is motivated by a combination of DESI BAO, high-$\ell$ CMB, and CMB
lensing measurements \cite{Sailer:2025lxj,2025ApJ...987L..29C}.
We refer to these as the Planck-like and
DESI-inspired fiducials, respectively.
The CMB parameter vector used in the forecasts is $\boldsymbol\Theta=(\omega_{\rm b},\omega_{\rm c},H_0,\tau,n_{\rm s},q,\Sigma m_\nu)$, where $\omega_i\equiv\Omega_i h^2$, $q\equiv\ln(10^{10}A_{\rm s})$, and we take the sum of the neutrino masses $\Sigma m_\nu=0.06\,\mathrm{eV}$. Since the SR relations depend on $\sigma_8$ rather than $A_{\rm s}$, we convert between the two with the symbolic-regression produced expression for $\sigma_8$ of Ref.~\cite{Bartlett:2023cyr}.

\begin{table}[t]
\centering
\small
\setlength{\tabcolsep}{5pt}
\renewcommand{\arraystretch}{1.10}
\caption{Fiducial parameters for the two forecasts. The value of $\tau$ is the one simulated with \texttt{21cmFAST}. The star-formation and escape-fraction normalizations and $M_{\rm turn}$ are given in logarithmic units. The X-ray inputs of the 21\,cm calculations are $\log_{10}L_X=40.5$, $E_0=500\,\mathrm{eV}$, and $\alpha_X=1$.}
\label{tab:fisher_fiducials}
\begin{tabular}{@{}lcc@{}}
\toprule
Parameter & Planck-like & DESI-inspired \\
\midrule
$\tau$ & 0.054 & 0.090 \\
$\Omega_{\rm b}h^2$ & 0.021231 & 0.021863 \\
$\Omega_{\rm c}h^2$ & 0.119478 & 0.122848 \\
$H_0\,[{\rm km\,s^{-1}\,Mpc^{-1}}]$ & 66.061 & 67.448 \\
$n_{\rm s}$ & 0.958133 & 0.967824 \\
$\sigma_8$ & 0.814691 & 0.823111 \\
$\ln(10^{10}A_{\rm s})$ & 3.02350 & 3.00289 \\
$\Sigma m_\nu\,[{\rm eV}]$ & 0.06 & 0.06 \\
\midrule
$\log_{10}f_{\star,10}$ & $-1.2098$ & $-1.0869$ \\
$\alpha_\star$ & 0.46188 & 0.43861 \\
$\log_{10}f_{\rm esc,10}$ & $-0.91609$ & $-0.55492$ \\
$\alpha_{\rm esc}$ & 0.30389 & $-0.35581$ \\
$\log_{10}(M_{\rm turn}/M_\odot)$ & 9.1046 & 8.3302 \\
$R_{\rm bubble}^{\rm max}\,[{\rm Mpc}]$ & 25.840 & 20.928 \\
\bottomrule
\end{tabular}
\end{table}

For the CMB we adopt a configuration inspired by the Simons Observatory (SO)~\cite{SimonsObservatory:2018koc}, with lensed power spectra computed using \texttt{CAMB}~\cite{Lewis:1999bs}. The Fisher matrix includes $TT$, $TE$ 
power spectra over $30\leq\ell\leq3000$, $EE$ power spectra over $2\leq\ell\leq5000$, and the lensing potential over $2\leq L\leq3000$, with $\ell$ being the angular multipoles. We use a sky fraction of 0.4 for the main survey and lensing and 0.1 for the low-$\ell$ polarization. The six frequency channels between 27 and $280\,\mathrm{GHz}$ are combined by inverse-noise weighting, with Gaussian beams and white-noise levels following the SO baseline~\cite{SimonsObservatory:2018koc}, and the lensing reconstruction noise is described by an analytic approximation. The spectra are differentiated by two-sided finite differences with steps $\max(0.02|\Theta_i|,10^{-4})$.

To include a prediction for the optical depth, we remove $\tau$ as a free CMB parameter. With ${\boldsymbol\phi}=(\omega_{\rm b},\omega_{\rm c},H_0,n_{\rm s},q,\Sigma m_\nu)$, a change in the remaining parameters changes $\tau$ by
\begin{equation}
\delta\tau=\sum_i a_i\,\delta\phi_i,\qquad
 a_i\equiv\left.\frac{{\rm d}\tau_{\rm SR}}{{\rm d}\phi_i}\right|_{\rm fid},
\label{eq:fisher_tau_linear}
\end{equation}
and the CMB Fisher matrix becomes~\cite{Liu:2015txa}
\begin{equation}
F^{\rm CMB+\tau}_{ij}=F^{\rm CMB}_{ij}
+a_i F^{\rm CMB}_{\tau j}+a_j F^{\rm CMB}_{i\tau}
+a_i a_j F^{\rm CMB}_{\tau\tau}.
\label{eq:fisher_tau}
\end{equation}
The slopes $a_i$ are total derivatives obtained by automatic differentiation of the SR expressions with \texttt{JAX}, including the implicit dependence of $\sigma_8$ on the other parameters. We use the three relations fitted to \texttt{21cmFAST} (Eq.~\ref{eq:tau_21cmf}), \texttt{AMBER} (Eq.~\ref{eq:tau_amber}, composed with Eqs.~\ref{eq:z_mid} to \ref{eq:z_asy}) and \texttt{Zeus21} (Eq.~\ref{eq:tau_zeus21}), and refer to each by the name of its simulation. After the substitution $\tau$ is a derived quantity, and its uncertainty follows from the reduced covariance $\mathbf{C} = (\mathbf{F}^{\rm CMB+\tau})^{-1}$ as $\sigma^2(\tau)= \mathbf{a}^{T}\mathbf{C} \mathbf{a}$.

In addition to simply asking how CMB constraints can be sharpened by reionization modeling, we also consider scenarios where additional probes are included. The first is the 21\,cm power spectrum $\Delta_{21}^2(k,z)$ computed with \texttt{21cmFAST}~\cite{Mesinger2011,Park:2018ljd} in $300\,\mathrm{Mpc}$ boxes with $128^3$ cells, in 19 redshift slices between $z=6$ and 15 and 30 logarithmic $k$ bins. Its derivatives come from central finite differences of simulations with respect to $(\omega_{\rm b},\omega_{\rm c},H_0,n_{\rm s},q)$, the six reionization parameters of Table~\ref{tab:fisher_fiducials}, and three X-ray parameters; $\Sigma m_\nu$ is not varied. The noise is computed with \texttt{py21cmsense}~\cite{Pober:2013jna,Pober:2012zz,Murray:2024the} for a HERA-like array~\cite{DeBoer:2016tnn} of 331 dishes of $14\,\mathrm{m}$ diameter, assuming 540 nights with six hours of observation, an $8\,\mathrm{MHz}$ bandwidth, receiver temperature $T_{\rm rcv}=100\,\mathrm{K}$, sky temperature $T_{\rm sky}=60\,\mathrm{K}(\nu/300\,\mathrm{MHz})^{-2.55}$, and a moderate foreground avoidance strategy of cutting out a ``wedge" in Fourier space~\cite{Murray:2024the}. Of the remaining Fourier modes, we keep those that fall within $0.05\leq k\leq1.5\,\mathrm{Mpc}^{-1}$ and write
\begin{equation}
F^{21}_{ij}=\sum_{z,k} \frac{1}{\sigma_{21}^2(k,z)}
\frac{\partial\Delta_{21}^2(k,z)}{\partial\phi_i}
\frac{\partial\Delta_{21}^2(k,z)}{\partial\phi_j},
\label{eq:fisher_21cm}
\end{equation}
where $\sigma_{21}$ includes thermal noise and sample variance, and correlations between redshift slices are neglected.

The second additional probe is a DESI-like galaxy survey~\cite{DESI:2016fyo} over $14,000\,\mathrm{deg}^2$ and $0.6\leq z\leq1.9$ in 13 bins. Emission-line galaxies, luminous red galaxies and quasars are combined into one effective tracer per bin, with bias normalizations $bD=0.84$, 1.7 and 1.2~\cite{DESI:2016fyo}, where $b$ is the bias of the tracer and $D$ is the linear growth factor. The redshift-space power spectrum is~\cite{Kaiser:1987qv}
\begin{equation}
P_g(k,\mu,z)=[b(z)+f(z)\mu^2]^2
e^{-(k\mu\sigma_v/H)^2}P_{\rm m}(k,z),
\label{eq:fisher_Pg}
\end{equation}
with $P_{\rm m}$ from \texttt{CAMB}, $f\simeq\Omega_{\rm m}(z)^{0.55}$~\cite{Linder:2005in} and $\sigma_v=300\,\mathrm{km\,s^{-1}}$. The Fisher matrix covers $0.01\leq k\leq0.25\,h\,\mathrm{Mpc}^{-1}$ with the effective-volume weighting of Ref.~\cite{Seo:2003pu}. Bias and redshift-space nuisance parameters are held fixed. When considering the CMB, $21\,\textrm{cm}$ power spectra, and galaxy surveys together, we add the three Fisher matrices under the assumption that the probes are independent, and marginalize over the nine astrophysical parameters.

Beyond considering different combinations of probes, we also pose the question of what happens if the reionization model used for data analysis is not the one that correctly describes the sky. Let $\tau_{\rm t}(\boldsymbol\phi)$ be the true relation between $\tau$ and the other parameters, and $\tau_{\rm a}(\boldsymbol\phi)$ the relation trusted in the analysis. If these differ at the fiducial parameter space point $\phi_0$ by $\Delta\tau_0=\tau_{\rm t}(\boldsymbol\phi_0)-\tau_{\rm a}(\boldsymbol\phi_0)$, to first order this offset changes the CMB spectra by $(\partial C_\ell/\partial\tau)\,\Delta\tau_0$, and the fit absorbs it by shifting the parameters by~\cite{Huterer:2005ez,Taylor:2006aw}
\begin{equation}
b_i=\sum_j \left(F^{-1}\right)_{ij}\left(F^{\rm CMB}_{j\tau}+a_jF^{\rm CMB}_{\tau\tau}\right)\Delta\tau_0,
\label{eq:fisher_bias}
\end{equation}
where $b_i=\hat\phi_i-\phi_{0,i}$ is the bias in parameter $\phi_i$, that is, the shift of its best-fit value $\hat\phi_i$ away from the fiducial value $\phi_{0,i}$ when the analysis uses the trusted relation. Here $\mathbf{F}$ is the Fisher matrix of the probe combination and $a_j$ are the slopes of the trusted relation.
The 21\,cm and galaxy data do not depend on $\tau$ explicitly and enter only through $F$. We quote each shift relative to the marginalized error, $b_i/\sigma_i$.

\subsection{Gain from the optical depth relation and modeling bias}
\label{subsec:fisher_results}

\begin{figure*}[t]
\centering
\includegraphics[width=0.78\textwidth]{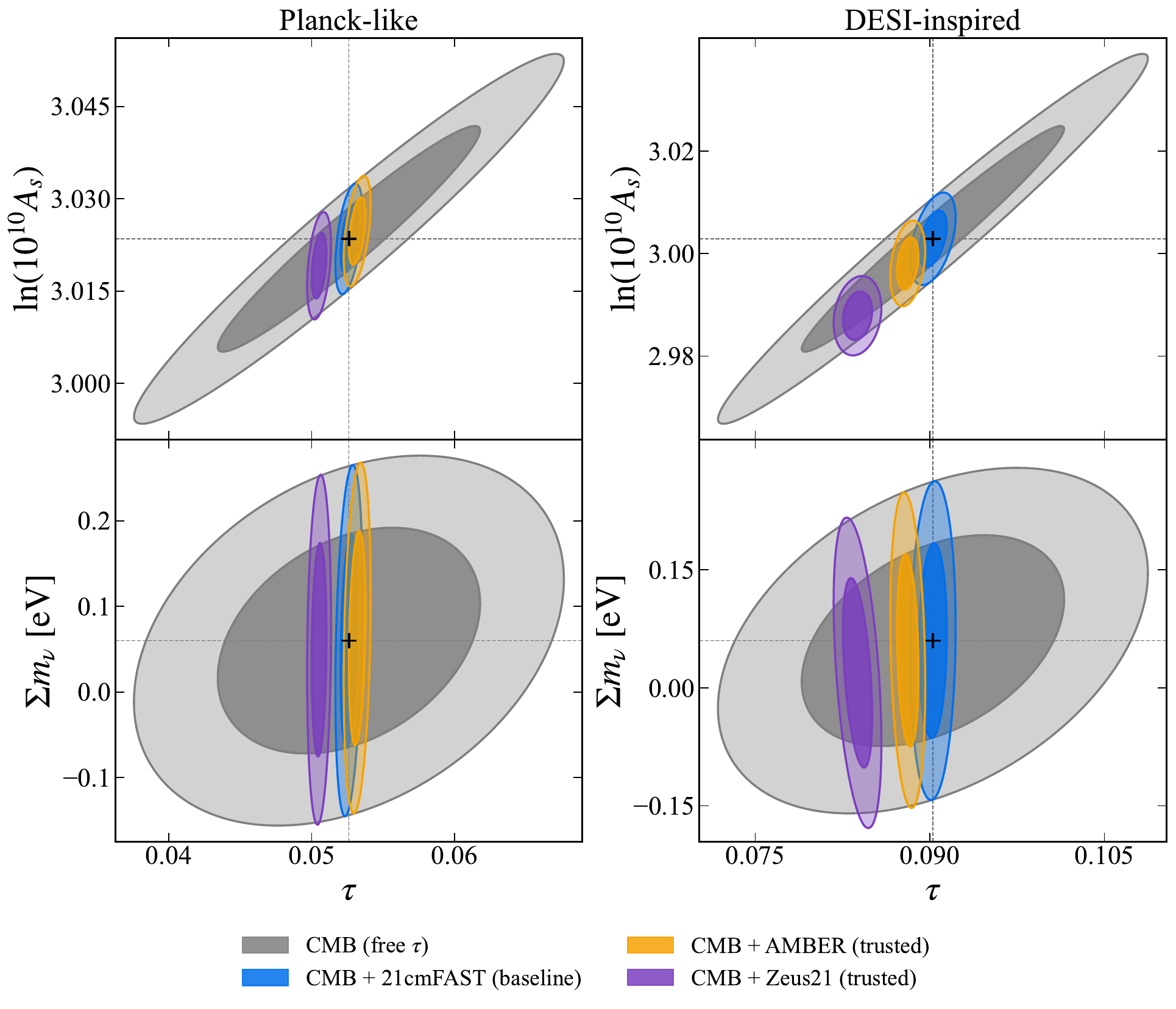}
\caption{Constraints on $\tau$ against $\ln(10^{10}A_{\rm s})$ (top) and $\Sigma m_\nu$ (bottom) at the Planck-like (left) and DESI-inspired (right) fiducials, with \texttt{21cmFAST} as the baseline relation. Grey contours show the CMB alone with free $\tau$. Coloured contours show the CMB with $\tau$ replaced by the relation fitted to \texttt{21cmFAST} (blue), \texttt{AMBER} (orange) or \texttt{Zeus21} (purple). The black plus marks the fiducial point, and the inner and outer contours show 68\% and 95\% credibility regions.}
\label{fig:fisher_bias}
\end{figure*}

To perform forecasts, we must select one of our three codes as a baseline. We arbitrarily select \texttt{21cmFAST} for this role in this paper, although we stress that this in no way endorses \texttt{21cmFAST} above \texttt{Zeus21} or \texttt{AMBER}. Using fiducial cosmological parameters and adjusting \texttt{21cmFAST} parameters appropriately gives us $\tau=0.0526$ for the Planck-like point and and $\tau = 0.0903$ for the DESI-inspired point. At the same points in parameter space the \texttt{AMBER} and \texttt{Zeus21} relations give 0.0532 and 0.0505 at the Planck-like point and 0.0881 and 0.0838 at the DESI-inspired point. Inserting these values into our Fisher machinery gives Fig.~\ref{fig:fisher_bias}, which shows the resulting constraints on parameters that are the most degenerate with $\tau$, and Table~\ref{tab:fisher_constraints_twofid} lists the marginalized errors.

\begin{table*}[t]
\centering
\small
\setlength{\tabcolsep}{4.4pt}
\renewcommand{\arraystretch}{1.10}
\caption{Marginalized $1\sigma$ forecast errors. Here $q=\ln(10^{10}A_{\rm s})$; errors in $q$ are in units of $10^{-3}$, those in $\Sigma m_\nu$ in meV, and those in $H_0$ in $\mathrm{km\,s^{-1}\,Mpc^{-1}}$. Rows labelled \texttt{21cmFAST}, \texttt{AMBER} and \texttt{Zeus21} replace $\tau$ as a free CMB parameter by the relation fitted to that simulation.}
\label{tab:fisher_constraints_twofid}
\begin{tabular}{@{}lrrrrrr@{}}
\toprule
& \multicolumn{3}{c}{Planck-like}
& \multicolumn{3}{c}{DESI-inspired}\\
\cmidrule(lr){2-4}\cmidrule(l){5-7}
Method & $10^3\sigma(q)$ & $\sigma(\Sigma m_\nu)$ & $\sigma(H_0)$
       & $10^3\sigma(q)$ & $\sigma(\Sigma m_\nu)$ & $\sigma(H_0)$\\
\midrule
CMB only & 12.09 & 86.94 & 1.338 & 14.51 & 88.30 & 1.323\\
\midrule
CMB $+$ \texttt{21cmFAST} & 3.61 & 82.54 & 1.325 & 3.63 & 81.53 & 1.303\\
CMB $+$ \texttt{AMBER} & 3.62 & 82.46 & 1.325 & 3.36 & 80.63 & 1.297\\
CMB $+$ \texttt{Zeus21} & 3.51 & 82.23 & 1.323 & 3.12 & 79.39 & 1.288\\
\midrule
CMB $+$ 21\,cm PS & 8.18 & 25.04 & 0.309 & 13.64 & 56.25 & 0.748\\
\midrule
CMB $+$ \texttt{21cmFAST} $+$ 21\,cm PS & 2.25 & 21.22 & 0.305 & 2.93 & 45.63 & 0.710\\
CMB $+$ \texttt{AMBER} $+$ 21\,cm PS & 2.30 & 21.24 & 0.305 & 2.77 & 44.89 & 0.706\\
CMB $+$ \texttt{Zeus21} $+$ 21\,cm PS & 2.23 & 21.15 & 0.305 & 2.75 & 44.40 & 0.701\\
\midrule
CMB $+$ \texttt{21cmFAST} $+$ 21\,cm $+$ DESI & 1.75 & 4.22 & 0.060 & 2.31 & 5.26 & 0.070\\
CMB $+$ \texttt{AMBER} $+$ 21\,cm $+$ DESI & 1.79 & 4.24 & 0.060 & 2.24 & 5.00 & 0.071\\
CMB $+$ \texttt{Zeus21} $+$ 21\,cm $+$ DESI & 1.74 & 4.18 & 0.060 & 2.35 & 5.23 & 0.070\\
\bottomrule
\end{tabular}
\end{table*}

Immediately apparent is the fact that modeling reionization removes most of the degeneracy between $A_{\rm s}$ and $\tau$. With the CMB alone, it reduces $\sigma(q)$ by a factor of 3.3 to 3.4 at the Planck-like point and 4.0 to 4.7 at the DESI-inspired point (Table~\ref{tab:fisher_constraints_twofid}).
The three models give similar errors, within 3\% of each other at the Planck-like point and 16\% at the DESI-inspired point. The neutrino mass constraints improve much less, between 5\% and 10\%, because $\Sigma m_\nu$ is degenerate with parameters that a $\tau({\boldsymbol \phi})$ relation does not fix.

The three different models do not predict the same $\tau$ even when at the same point in parameter space, and this offset shifts the contours along the CMB degeneracy direction. Table~\ref{tab:fisher_bias} lists the resulting shifts when \texttt{21cmFAST} is used as the baseline, in units of the error obtained with the model trusted during data analysis. At the Planck-like point, trusting \texttt{AMBER} ($\Delta\tau_0=-5.8\times10^{-4}$) moves $q$ by $+0.35\sigma$,
while trusting \texttt{Zeus21} ($\Delta\tau_0=+2.1\times10^{-3}$) moves it by $-1.25\sigma$, or $-4.4\times10^{-3}$. At the DESI-inspired point the offsets are larger, $2.2\times10^{-3}$ for \texttt{AMBER} and $6.5\times10^{-3}$ for \texttt{Zeus21}, and the shifts in $q$ reach $-1.4\sigma$ and $-4.8\sigma$. In all cases the shift in $\Sigma m_\nu$ is a small fraction of its error. Choosing \texttt{AMBER} or \texttt{Zeus21} as the truth gives the same picture. The shifts change sign when truth and trusted models are swapped; trusting \texttt{AMBER} when \texttt{Zeus21} is true shifts $q$ by $5.75\times10^{-3}$, close to the $5.67\times10^{-3}$ obtained by going through \texttt{21cmFAST}. A different choice of truth therefore only moves the origin of Fig.~\ref{fig:fisher_bias}.

\begin{table}[t]
\centering
\small
\setlength{\tabcolsep}{4pt}
\renewcommand{\arraystretch}{1.10}
\caption{Parameter shifts from Eq.~\eqref{eq:fisher_bias} with \texttt{21cmFAST} as the baseline relation and \texttt{AMBER} or \texttt{Zeus21} as the trusted relation in the analysis. Shifts are in units of the marginalized error of the trusted analysis.}
\label{tab:fisher_bias}
\begin{tabular}{@{}lrrrr@{}}
\toprule
& \multicolumn{2}{c}{Planck-like} & \multicolumn{2}{c}{DESI-inspired}\\
\cmidrule(lr){2-3}\cmidrule(l){4-5}
Trusted & $b_q/\sigma$ & $b_{m_\nu}/\sigma$ & $b_q/\sigma$ & $b_{m_\nu}/\sigma$\\
\midrule
\multicolumn{5}{@{}l}{\textit{CMB with the $\tau$ relation}}\\
\texttt{AMBER}  & $+0.35$ & $+0.04$ & $-1.43$ & $-0.15$\\
\texttt{Zeus21} & $-1.25$ & $-0.13$ & $-4.81$ & $-0.52$\\
\midrule
\multicolumn{5}{@{}l}{\textit{$+$ 21\,cm}}\\
\texttt{AMBER}  & $+0.53$ & $+0.10$ & $-1.73$ & $-0.27$\\
\texttt{Zeus21} & $-1.92$ & $-0.35$ & $-5.42$ & $-0.87$\\
\midrule
\multicolumn{5}{@{}l}{\textit{$+$ 21\,cm $+$ DESI}}\\
\texttt{AMBER}  & $+0.63$ & $+0.38$ & $-1.93$ & $-1.31$\\
\texttt{Zeus21} & $-2.28$ & $-1.35$ & $-5.75$ & $-3.86$\\
\bottomrule
\end{tabular}
\end{table}

Before reading these shifts as disagreements between the codes, we qualitatively compare the offsets with the accuracy of the fits. At the Planck-like point the \texttt{AMBER} offset equals the typical difference between the \texttt{AMBER} and \texttt{21cmFAST} relations (Sec.~\ref{sec:simulation_comparison}), while the \texttt{Zeus21} offset is almost three times higher than the combined RMSE of the two fits. At the DESI-inspired point, \texttt{Zeus21} offset exceeds the error scales. The \texttt{Zeus21} offset is possibly reflecting a real difference between the codes, mild at the Planck-like point and very apparent at the DESI-inspired point.

\subsection{Adding 21\,cm and galaxy clustering information}
\label{subsec:fisher_additional_probes}

\begin{figure*}[t]
\centering
\includegraphics[width=0.78\textwidth]{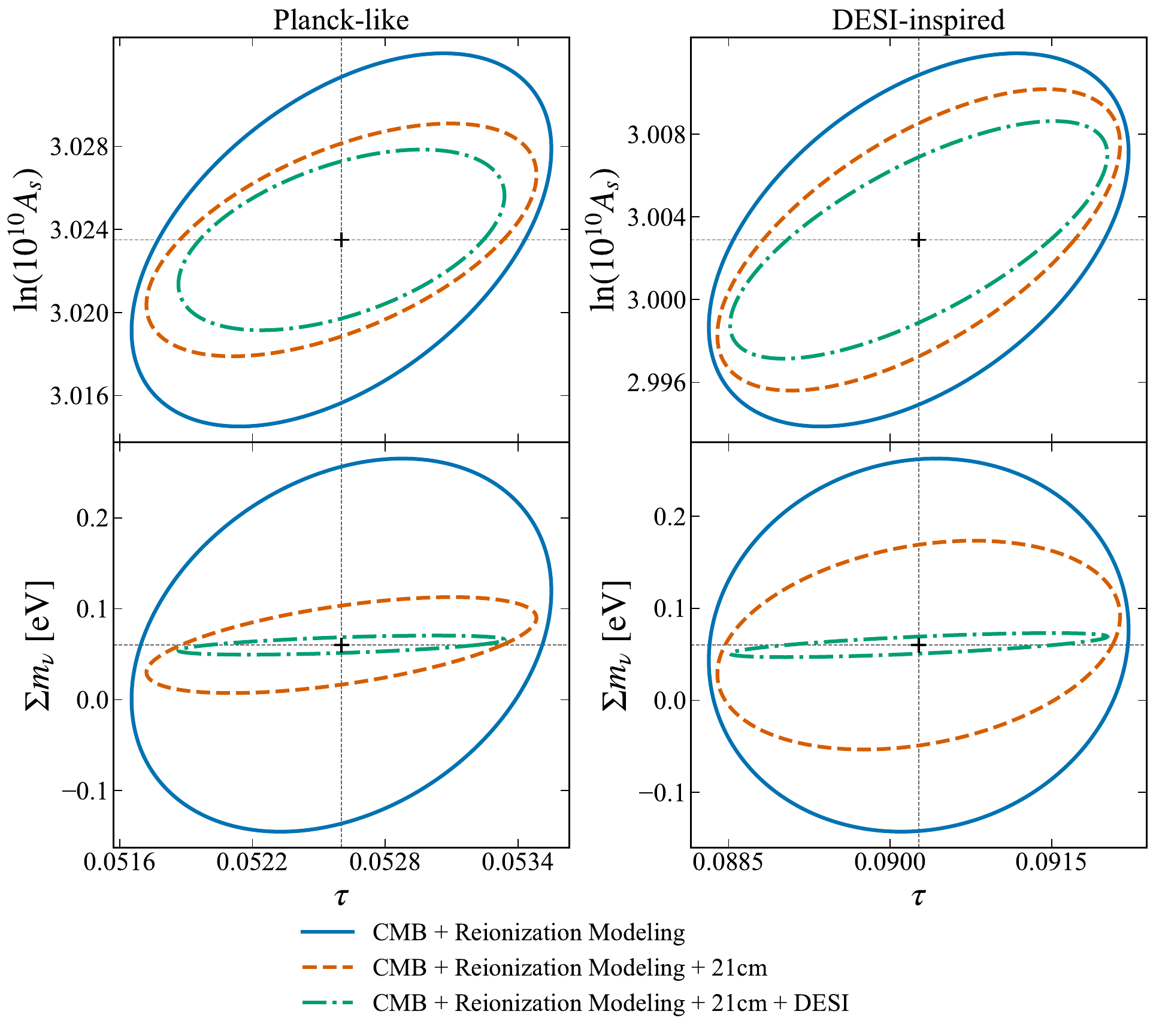}
\caption{Constraints at the Planck-like (left) and DESI-inspired (right) fiducials using the \texttt{21cmFAST} relation. Blue contours show the CMB with the optical-depth relation, red adds the HERA-like 21\,cm power spectrum, and green also adds DESI-like galaxy clustering. The $\tau$ uncertainty is obtained by propagation through the relation. All contours show 95\% credibility regions.}
\label{fig:fisher_probes}
\end{figure*}

Fig.~\ref{fig:fisher_probes} shows how the constraints tighten as the other probes are added, using \texttt{21cmFAST} both as the baseline and the trusted model. The variances from assuming other models as the trusted analysis models are similar, and the results differ only by the bias offsets listed in Table~\ref{tab:fisher_bias}.

The 21\,cm power spectrum helps break the CMB degeneracies involving $H_0$, $\omega_{\rm c}$ and $\Sigma m_\nu$.  At the Planck-like point, $\sigma(\Sigma m_\nu)$ falls from 82.5 to $21.2\,\mathrm{meV}$ and $\sigma(q)$ from 3.6 to $2.3\times10^{-3}$. The gain is smaller at the DESI-inspired point, where $\sigma(\Sigma m_\nu)$ reaches $45.6\,\mathrm{meV}$, reflecting the different 21\,cm sensitivity in the two reionization scenarios. The galaxy survey then adds the scale-dependent suppression of clustering by neutrino free streaming, which separates $\Sigma m_\nu$ from the primordial amplitude. In the full combination $\sigma(\Sigma m_\nu)=4.2$ and $5.3\,\mathrm{meV}$ at the two fiducials, about 20 and 17 times better than the CMB alone, and $\sigma(H_0)$ improves by about 22 and 19.

As the errors shrink, the modeling bias becomes more important. Because the other probes do not depend on $\tau$ explicitly, the physical shifts barely change as they are added, but they grow relative to the shrinking errors (Table~\ref{tab:fisher_bias}). Trusting \texttt{Zeus21} at the Planck-like point shifts $\Sigma m_\nu$ by $-0.13\sigma$ with the CMB alone and by $-1.35\sigma$ in the full combination, and shifts $q$ by $-2.3\sigma$. At this fiducial, the \texttt{Zeus21} offset is of $2\times10^{-3}$ in $\tau$, which exceeds the fitting errors and reflects a difference between the simulations. It is thus enough to bias $q$ by more than $2\sigma$ once all probes are combined. At the DESI-inspired point, where the offset is larger, the full combination gives shifts of $-3.9\sigma$ in $\Sigma m_\nu$, $-5.7\sigma$ in $q$, and $+3.2\sigma$ in $\omega_{\rm c}$.

These forecasts use an approximate lensing reconstruction noise, neglect correlations between 21\,cm redshift slices, and fix the galaxy nuisance parameters, so the absolute errors are optimistic. The comparison between the relations does not rely on these choices. Our results show that a model for $\tau$ substantially tightens the constraint on $A_{\rm s}$ and, combined with 21\,cm and galaxy data, contributes to a precise measurement of $\Sigma m_\nu$. At that precision, however, differences of order $10^{-3}$ in the predicted $\tau$ produce shifts comparable to the statistical errors. Such differences already arise between reionization codes, so the choice of simulation must be treated as a source of systematic error in forecasts of this kind.

\section{Summary and discussion}
\label{sec:summary_discussion}

In this work, we have used symbolic regression to describe the CMB optical depth, $\tau$, in terms of the cosmological and astrophysical parameters of three reionization simulations: \texttt{21cmFAST}, \texttt{AMBER}, and \texttt{Zeus21}. Our goal was not only to replace repeated simulation runs with faster predictions. We also obtained explicit relations that could be differentiated, examined for parameter dependencies, and compared across codes with different descriptions of reionization. As a worked example, we ran a series of Fisher matrix forecasts to elucidate how reionization models can improve CMB constraints, paying particular emphasis to the model dependence of our conclusions.

In general, we are able to produce symbolic regression fits with small errors across wide ranges in parameter space. For the three simulations used in this paper (\texttt{21cmFAST}, \texttt{AMBER}, and \texttt{Zeus21}), we were able to achieve accuracies that were small compared to current observational uncertainties and comparable to those produced by a neural network. Where our approach shows additional promise beyond a black box emulator is in interpretability. Several features of the fitted equations have direct physical interpretation. All three contain the expected $\Omega_{\rm b}$ prefactor, although we imposed this structure in the \texttt{21cmFAST} and \texttt{Zeus21} fits after we saw that it was very often picked up by the symbolic regression package. The repeated appearance of combinations involving $f_{\star,10}$ and $f_{\rm esc,10}$, or $\alpha_\star$ and $\alpha_{\rm esc}$, is also consistent with the dependence of reionization on the supply of ionizing photons and the fraction that escapes into the intergalactic medium. Note that these combinations reveal degeneracies in the prediction of $\tau$; they do not imply that star formation and photon escape have identical effects on every reionization observable. In particular, an integrated optical depth cannot retain all the information contained in the redshift evolution and spatial distribution of ionized regions.

Our symbolic regression expressions give useful information beyond the value of $\tau$. Their first derivatives generally follow the simulation derivatives for parameters with the strongest effect on the optical depth. Agreement is less uniform for weakly contributing parameters in \texttt{21cmFAST} and \texttt{Zeus21}. Thus, a small error in $\tau$ does not by itself guarantee a similarly small error in every derivative. This matters for Fisher forecasts and other calculations that depend on local parameter responses. It is an advantage of the symbolic expressions that these responses can be evaluated at many points without additional reionization simulations, while their accuracy can still be checked against finite differences where necessary.

A further result is that the simulation outputs can be compared after bringing their input parameters into a common form. We used symbolic regression to relate the \texttt{AMBER} reionization history parameters to the inputs of \texttt{21cmFAST} and \texttt{Zeus21}, allowing the \texttt{AMBER} optical-depth relation to be evaluated and differentiated in the same parameter basis. After bringing all three codes onto a common set of parameters for a fair comparison, we found that they broadly agreed on the relative importance of each parameter's influence on $\tau$.

Our newfound ability to perform like-for-like comparisons between models was then practiced in a series of Fisher forecasts. We examined how the incorporation of a reionization model could improve CMB constraints. In particular, replacing $\tau$ as an independent CMB parameter with a model relation helps break its degeneracy with the primordial amplitude $A_{\rm s}$. For the CMB alone, this reduces the uncertainty in $q\equiv\ln(10^{10}A_{\rm s})$ by factors of a few across different fiducial $\tau$ values.
All three models give similar forecast uncertainties, indicating that the improvement is not strongly dependent on the choice of reionization code.

However, similar forecast uncertainties do not necessarily imply agreement on the inferred parameter values. In other words, while the forecasted variances may be similar, we do see biases in the inferences when there is a mismatch between the ``true" baseline model used for simulation and the ``trusted" model used for analysis. Using \texttt{21cmFAST} as a baseline and fitting with \texttt{AMBER} produces biases that are comparable to the accuracy of the symbolic regression fits. In contrast, with \texttt{Zeus21} the biases are more significant, with best-fit $q$ values for instance shifting by approximately $1.3\sigma$ and $4.8\sigma$ for the low $\tau$ and high $\tau$ scenarios, respectively.\footnote{We reemphasize that this is not a statement regarding the correctness of any code. For example, had we chosen \texttt{Zeus21} as our baseline model our discussion would have been phrased in terms of biases in using \texttt{21cmFAST} or \texttt{AMBER} in data analysis.} Note that here $\sigma$ refers to the reduced error \emph{after} the inclusion of a reionization model, and thus the bias is still small compared to the uncertainty due to the original parameter degeneracies. This suggests that even amidst modeling uncertainties, it is generally worthwhile to consider reionization modeling when performing CMB parameter inferences, although of course this conclusion may change when more models are considered beyond the three in this paper.

Adding $21\,\textrm{cm}$ power spectrum measurements and galaxy surveys further improves constraints on cosmological parameters. A HERA-like 21\,cm power spectrum helps break degeneracies between cosmological and astrophysical parameters, with a larger improvement at the Planck-like fiducial $\tau$. The galaxy power spectrum measured at DESI-level errors provides further complementary information. With all three probes combined, the uncertainty in $\Sigma m_\nu$ reaches approximately $\sim 5\,{\rm meV}$. However, as the statistical uncertainties decrease, differences between the optical depth relations become more significant. For example, when the \texttt{Zeus21} relation is used in analysis instead of the \texttt{21cmFAST} baseline, the shift in $\Sigma m_\nu$ increases from $0.13\sigma$ with the CMB alone to $1.35\sigma$ in the full combination at the Planck-like fiducial, and reaches $3.9\sigma$ at the DESI-inspired fiducial $\tau$. This illustrates how reducing model uncertainties in reionization is a moving target, with theory requirements becoming more and more stringent as observations become more precise.

Our forecasts should be regarded as illustrative. We use an approximate CMB lensing reconstruction noise, neglect possible correlations between the 21\,cm redshift slices, and do not vary galaxy-bias and redshift-space nuisance parameters. These simplifications can make the forecast constraints optimistic, particularly for $A_{\rm s}$ and $\Sigma m_\nu$. Nevertheless, the analysis provides a useful comparison of the three optical-depth relations within a common forecast setup. The precise significance of the parameter shifts will depend on the assumptions made about the surveys and their uncertainties.

In this paper we used the optical depth as a proof-of-concept study for applying symbolic regression to reionization studies. The main value of symbolic regression in this setting is that it does not simply give a fast approximation to a simulation output. It provides an explicit and differentiable link between input physics and an observable, making it possible to inspect parameter combinations, place different simulations in a common language, and to efficiently test their consequences for cosmological inference. However, symbolic regression is a promising technique that could also be applied to other quantities, such as the 21\,cm power spectrum. Future work along these lines would test aspects of reionization that are lost when its history is compressed into $\tau$. Symbolic regression could therefore play a key role in unlocking our ability to routinely include well-modeled reionization physics in cosmological analyses. While challenging, this is a potentially rewarding prospect that would take cosmological parameter constraints to the next level for both existing and future datasets.

\section*{Acknowledgments}
The authors would like to acknowledge Andrei Li, Robert Pascua, Hannah Fronenberg, Jade Ducharme, Joshua Goodeve, Kim Morel, Laurie Amen, Maria Bayder, Mattéo Blamart, Michael Wilensky, Patrick Horlaville, Sophia Rubens, Hy Trac, Julian Muñoz, Yacine Ali-Haïmoud, Rémi Tellier, and Joris Theil for the useful discussions which contributed to this work. M.D.V. acknowledges support from the Trottier Space Institute through a Summer Undergraduate Research Award. A.L. acknowledges support from the Natural Sciences and Engineering Research Council of Canada through the Discovery Grants Program and the Alliance International Program, as well as from the William Dawson Scholar Program at McGill University. 
D.S. acknowledges the support of the Canada $150$ Chairs program, the Fonds de recherche du Qu\'{e}bec Nature et Technologies (FRQNT) and Natural Sciences and Engineering Research Council of Canada (NSERC) joint NOVA grant, and the Trottier Space Institute Postdoctoral Fellowship program.
This research was enabled in part by support provided by Calcul Québec (\href{https://www.calculquebec.ca/}{https://www.calculquebec.ca/}) and the Digital Research Alliance of Canada (\href{https://www.alliancecan.ca}{https://www.alliancecan.ca}). 
This work was also enabled by the Python packages \texttt{numpy} \cite{numpy}, \texttt{pandas} \cite{reback2020pandas, mckinney-proc-scipy-2010}, \texttt{matplotlib} \cite{Hunter:2007}, \texttt{sympy} \cite{sympy}, \texttt{pytorch} \cite{Paszke2019}, \texttt{PySR} \cite{Cranmer2023}, and \texttt{JAX} \cite{jax2018github}.

\bibliography{apssamp}

\end{document}